\documentclass[useAMS,usenatbib,usegraphicx,a4]{mn2e}
\usepackage{times}
\usepackage{url}
\usepackage{epsfig}
\usepackage[flushleft]{threeparttable}
\usepackage{amssymb}
\usepackage{amsmath}
\usepackage{pdfpages}
\usepackage{todonotes}
\usepackage{array}
\usepackage{comment}

\def\arcsec{^{\prime\prime}}

\def\apj{ApJ}

\def\mnras{MNRAS}

\def\aj{AJ}
\def\aap{A\&A}

\def\pasp{PASP}

\newcommand{\refbf}{} 

\title[Ultra-luminous quasars from SkyMapper]{Ultra-luminous quasars at redshift $z>4.5$ from SkyMapper}

\author[C. Wolf et al.]{Christian Wolf$^{1,2}$, Wei Jeat Hon$^3$, Fuyan Bian$^{4}$, Christopher A. Onken$^{1,2}$, Noura Alonzi$^{3,5}$, 
\newauthor Michael A. Bessell$^1$, Zefeng Li$^{1,2}$, Brian P. Schmidt$^{1,2}$, Patrick Tisserand$^6$ \\ 
$^1$Research School of Astronomy and Astrophysics, Australian National University, Canberra ACT 2611, Australia, E-mail: christian.wolf@anu.edu.au\\
$^2$ARC Centre of Excellence for All-sky Astrophysics (CAASTRO) \\
$^3$School of Physics, University of Melbourne, Parkville, Victoria 3010, Australia \\
$^4$European Southern Observatory, Alonso de C{\'o}rdova, Vitacura, Chile \\
$^5$Department of Physics and Astronomy, King Saud University, Riyadh 11451, Saudi Arabia\\
$^6$Sorbonne Universit\'{e}s, UPMC Univ Paris 6 et CNRS, Institut d'Astrophysique de Paris, 98 bis bd Arago, F-75014 Paris, France \\
}

\begin{document}

\date{draft \today}
\maketitle


\begin{abstract}
The most luminous quasars at high redshift harbour the fastest-growing and most massive black holes in the early Universe. They are exceedingly rare and hard to find. Here, we present our search for the most luminous quasars in the redshift range from $z=4.5$ to $5$ using data from SkyMapper, {\it Gaia} and {\it WISE}. We use colours to select likely high-redshift quasars and reduce the stellar contamination of the candidate set with parallax and proper motion data. In $\sim$12,500~deg$^2$ of Southern sky, we find 92 candidates brighter than $R_p=18.2$. Spectroscopic follow-up has revealed {\refbf 21 quasars at $z\ge 4$ (16 of which are within $z=[4.5,5]$), as well as several} red quasars, BAL quasars and objects with unusual spectra, which we tentatively label {\refbf OFeLoBALQSOs} at redshifts {\refbf of} $z\approx 1$ to $2$. This work {\refbf lifts} the number of known bright $z\ge 4.5$ quasars in the Southern hemisphere {\refbf from 10 to 26} and brings the total number of quasars known at $R_p<18.2$ and $z\ge 4.5$ to {\refbf 42}. 
\end{abstract}

\section{Introduction}\label{intro}

Quasars at high redshift point to the most extreme supermassive black holes (SMBHs) in the early Universe. They are centres of early galaxy and star formation and provide the first evidence for heavy chemical elements not created by Big Bang nucleosynthesis. They are interesting objects in their own right as they showcase the existence of extreme black holes, whose formation we struggle to understand \citep[][]{BrommLoeb03, Volonteri12, Pacucci15}. They are beacons of high luminosity that shine through assemblies of matter, which are otherwise virtually impossible to detect, let alone characterise; through that they support studies of the buildup of chemical elements and early structure in the universe \citep[e.g.][]{ Ryan-Weber09, Simcoe11, Diaz14}.

Traditionally, high-redshift quasars have been hard to find, as they are very rare. Unsurprisingly, the field was propelled from an aspirational undertaking into a mainstream science by the advent of the first truly wide-area deep  multi-band survey, the Sloan Digital Sky Survey \citep[SDSS;][]{SDSS}, as evidenced by a series of works from \citet{Fan01} to \citet{Jiang16}. One frontier of this work is to push searches to higher redshifts \citep{Mortlock11, Banados18}, while the other is to push to the highest and rarest luminosities \citep{Wu15,Wolf18b}.

High-redshift quasars are proverbial needles in the haystack, even more so as we look for the rarest, most luminous specimens. Quasars at high redshift appear star-like and their colours are to some extent similar to those of cool stars in the Milky Way. True quasars are vastly outnumbered by cool stars, and as we move towards the highest and most interesting potential luminosities, the fraction of true quasars in any candidate sample reaches zero.

Searches for luminous high-redshift quasars have identified samples of them in parts of colour space, where contamination by stars is modest. E.g., \citet{Wang16} used SDSS, which covers a large fraction of the Northern sky, in combination with data from the all-sky Widefield Infrared Survey Explorer \citep[{\it WISE};][]{Wright10} and found {\refbf 72 new quasars at redshifts $z>4.5$, bringing their total number known at the time to 795}. 

Further work has ventured into the regions in redshift where contamination by stars is a serious problem, finding a smaller number of objects despite strong efforts \citep{Yang17,Wang17}. Many searches have also added near-infrared data from the UKIRT Deep Infrared Sky Survey \citep[UKIDSS;][]{Warren07}, the UKIRT Hemisphere Survey \citep[UHS;][]{Dye18}, and the VISTA Hemisphere Survey \citep[VHS;][]{VHS} to improve the selection at various redshifts \citep{Schindler18,Schindler19a,Schindler19b,Yang19a,Yang19b}.

At all redshifts the luminosity function of quasars is known to be very steep such that quasars of very high luminosity are extremely rare. E.g., {\refbf among 1054} quasars at redshift $z\ge 4.5$ {\refbf listed in the MILLIQUAS catalogue \citep[version 6.2 from May 2019]{Flesch15}, only 18} are brighter than $R_p=18$ mag {\refbf as measured by the {\it Gaia} mission \citep{GaiaDR2}} of the European Space Agency (ESA). {\refbf By now,} the {\it Gaia} mission has provided a true all-sky survey with homogeneous calibration in the optical wavelength regime. Among the very brightest objects, where every candidate is a priori expected to be a star, there are still hidden surprises such as the ultra-luminous quasar J0010+2802, which harbours a black hole of 12 billion solar masses \citep{Wu15}. 

The Southern sky has only recently seen the publication of a hemispheric digital survey, i.e. the SkyMapper Southern Survey \citep{Wolf18a,Onken19}. By combining SkyMapper with {\it WISE}, a similar search for high-redshift quasars has become possible {\refbf across the whole Southern sky}. This search has also revealed a surprise, in the form of the intrinsically most luminous quasar found so far \citep[J2157-3602;][]{Wolf18b}. As the object appears not to be gravitationally lensed, it is expected to harbour the fastest-growing black hole we currently know.

This paper presents our first stage of searching for high-{\refbf z} quasars in the Southern sky. Besides SkyMapper and {\it WISE}, we use data from the {\it Gaia} satellite, which measures parallaxes and proper motions for the more nearby stars in our Milky Way and thus identifies by far most of the common cool dwarf stars that may look similar to high-redshift quasars. Our search here aims at the most luminous quasars we can find, where the previously staggering contamination by stars made spectroscopic confirmation unaffordable. Owing to the {\refbf release}
of {\it Gaia} DR2 \citep{GaiaDR2}, this section of parameter space can now be tackled with confidence.

In Section~\ref{data_methods}, we describe data and methods used to select our candidate sample and in Section~\ref{results} our spectroscopic {\refbf results}, before we close with a summary. We adopt a flat $\Lambda$CDM cosmology with $\Omega_\Lambda=0.7$ and $H_0=70$~km~s$^{-1}$~Mpc$^{-1}$. We use Vega magnitudes for {\it Gaia} and {\refbf infrared} data and AB magnitudes for SkyMapper passbands {\refbf (where $i_{\rm AB}-i_{\rm Vega}=0.40$ and $z_{\rm AB}-z_{\rm Vega}=0.52$)}.

\begin{figure}
\begin{center}
\includegraphics[angle=270,width=0.9\columnwidth,clip=true]{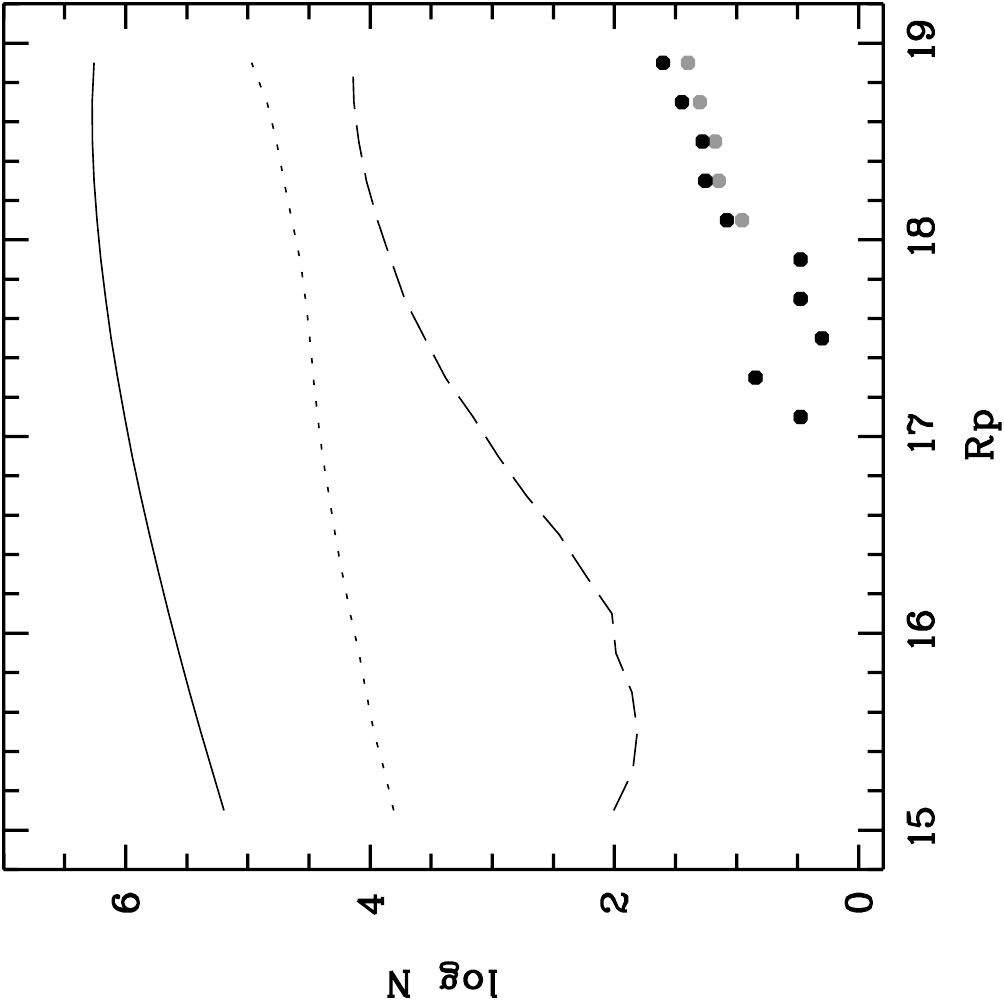} 
\caption{Differential number counts of objects with red {\it Gaia}/ {\it WISE} colours in the range covered by $z>4.5$ quasars. Solid line: all objects satisfying criteria specified in Sect.~\ref{data_gaia}. Dotted line: sample after removing objects with parallaxes or proper motions identifying them as Milky Way stars. Dashed line: the sample after removing objects without parallax and proper motion measurements. Symbols: known high-redshift quasars before (black) and after (grey) removing those excluded by the dashed-line rule.
\label{NCs}}
\end{center}
\end{figure}

\section{Data}\label{data_methods}

In this section, we describe how we select candidates for high-redshift quasars in the Southern hemisphere from survey catalogues using information on spectral energy distributions (SEDs) and on apparent motions that can be seen for nearby stars. Quasars at high redshift are characterised by three features in their SEDs that help differentiate them from other objects: 
\begin{enumerate}
    \item At wavelengths shorter than the Ly~$\alpha$ emission line the intrinsic quasar light is strongly attenuated by intergalactic absorption, creating a red colour between passbands located on either side of the line (e.g. at $z\approx 4.8$ the Ly~$\alpha$ line appears at 700~nm).
    \item At wavelengths {\refbf just} longer than the Ly~$\alpha$ emission line the typical quasar continuum is bluer than most other objects.
    \item Relative to stars, the optical-minus-mid-infrared colours of high-redshift quasars are redder. {\refbf Colours such as $z-W1$ capture mostly the blue Rayleigh-Jeans tail in the spectra of cool stars, while the accretion disk continuum in quasar SEDs comprises black-body contributions across a range in temperature, and given the high redshift we do not see their tails}.
\end{enumerate}

Following the example of \citet{Wang16}, we use optical as well as mid-infrared photometry to select quasar candidates. Mid-infrared photometry from {\it WISE} has been immensely helpful for selecting quasars with high completeness across the whole range of redshifts from $z\approx 0$ to beyond $z\approx 4$ \citep[e.g.][]{Wu12,Paris18}. It is just at redshifts $z>4$ that the contamination by cool stars has traditionally been a challenge,
which is here addressed with data from {\it Gaia}.

\subsection{Parallax and proper motion information from {\it Gaia}}\label{data_gaia}

The Data Release 2 (DR2) of the {\it Gaia} mission \citep{GaiaDR2} is the first and currently only release with parallax and proper motion measurements for objects as faint as we are targeting in this work. The release also contains photometry in two very broad passbands that cover roughly the wavelength range from 330 to 660~nm ($B_p$) and 630 to 1000~nm ($R_p$), respectively.

{\refbf How useful parallaxes are for differentiating Galactic and extragalactic sources depends on the colours of the selected sources, hence it is not useful to consider a random subsample of {\it Gaia} sources. Instead we} consult the sample of {\refbf 795 quasars at $z>4.5$} collated by \citet{Wang16} to learn what range of {\it Gaia}/ {\it WISE} colours they occupy and to quantify how much {\it Gaia} helps to reduce the contamination of candidate samples {\refbf for high-z quasars specifically. The vast majority of known quasars at $z>4.5$ {\refbf are} fainter than 19 mag in SDSS $i$/$z$ bands. Some are too faint to be detected by {\it Gaia}, and at the highest redshifts we expect that their dominant flux is redshifted beyond the $R_p$ passband. However, the 111 brightest objects in the \citet{Wang16} compilation, with either $i_{\rm AB}<19.2$ or $z_{\rm AB}<18.8$, all have a counterpart in {\it Gaia} DR2. Six of those 111 have $G$ band but no $B_p$ and $R_p$ photometry, while only one of them is bright enough to be missing from our later selection due to this issue.}

{\refbf At this point}, we use only simple colour selections, as the {\it Gaia} argument does not depend on details. The detailed colour selection for this first candidate list among our works on high-redshift quasars will only be finalised in Sect.~\ref{coloursel}. \citet{Wang16} show that $z>4.5$ quasars cover the interval $W1-W2=[0,2]$, although their own search is constrained to $W1-W2>0.5$ to avoid the dramatic stellar contamination below this threshold; this argument is based both on empirically found quasars as well as on the colour tracks of simulated quasars. They use further cuts, such as red colours in $z_{\rm SDSS}-W1$, for their candidates, and here we like to replace the SDSS filters with {\it Gaia} photometry as the latter is available for the whole sky. {\refbf Using dereddened colours (see Sect.~\ref{coloursel} for details),} we find that the selection $(B_p-R_p)_0>1.8$, $(R_p-W1)_0>2$, $W1-W2>0$ retains {\refbf all of their $z>4.5$ quasars at $R_p<19$ except for one at $R_p=18.99$}. We further select objects at galactic latitude $|b|>20\degr$ to avoid the denser star fields of the Galaxy, leaving a sky area of $\sim 26\,000$~deg$^2$. In Fig.~\ref{NCs} we show the number counts of {\it Gaia} objects satisfying these criteria: the solid line represents the whole sample of objects satisfying the simple colour cuts above. At $R_p<18$, the full object sample contains $\sim 11$ million objects. The symbols represent quasars from MILLIQUAS v6.2 \citep{Flesch15}, which at $R_p<18$ contains only 18 objects. Down to $R_p<19$, the numbers are 135 known quasars out of $\sim 20$ million objects in total.

We then remove objects with parallaxes or proper motions suggestive of Milky Way stars from this sample. At $R_p\la 18$, {\it Gaia} measures parallaxes with an error of $\sim 0.35$ milli-arcsec (mas) per year and proper motions with $\sim 0.5$ milli-arcsec (mas) per year in each coordinate axis. We thus choose to remove objects with a parallax of $>1$~mas per year and those with a proper motion of $\sqrt{{\rm pmra}^2+{\rm pmdec}^2}>\sqrt{2}$~mas per year. At $R_p<18$ this leaves {\refbf $\sim$300,000 objects}, whose number counts are shown as the dotted line. Hence, 97.3\% of the initial sample have been removed based on positive evidence of star-like motions.

An optional, more stringent, step is to remove objects for which {\it Gaia} does not provide any parallax and proper motion measurements in DR2. This will produce a purer candidate list at the risk of introducing incompleteness. Adding this criterion shrinks the sample to only {\refbf 21,923} objects at $R_p<18$ or 1 part in {\refbf $\sim$500} of the original sample (dashed line in Fig.~\ref{NCs}). 

Using {\refbf MILLIQUAS} we estimate the incompleteness caused by applying cuts based on {\it Gaia}'s motion data, which is {\refbf appropriate} since the previously known quasars were found before {\it Gaia} data had become available, and used only colour criteria. At $R_p<18$ all 18 known quasars pass the {\refbf motion} cuts, but incompleteness sets in {\refbf when going fainter}, mostly because the errors on {\it Gaia}'s proper motion measurements increase significantly. Fig.~\ref{NCs} shows the known quasars passing the stringent motion-based cuts as grey symbols. At $R_p=[18,18.5]$ and $R_p=[18.5,19]$, the cuts lose {\refbf $\sim$20\%} and {\refbf $\sim$33\% of the} quasars, respectively. Of course, a more complete and efficient selection function could take motion errors into account on a per-object basis, which will be {\refbf advisable when exploring fainter magnitudes}, but as this is beyond the scope of this paper, we continue with the selection motivated above. 

We summarise here that, before further colour selections are made, we reduced a list of {\refbf  $\sim$11}~million objects at $R_p=[15,18]$ to {\refbf $\sim$22,000} objects based on {\it Gaia}'s motion data. This list covers 26,000~deg$^2$ of sky, in which 18 high-redshift quasars had been {\refbf listed in MILLIQUAS}, although not the whole area had been searched systematically before SkyMapper data became available.

\subsection{The SkyMapper Southern Survey} 

SkyMapper is a 1.3m-telescope at Siding Spring Observatory near Coonabarabran in New South Wales, Australia \citep{Keller07}. It started a digital survey of the whole Southern hemisphere in 2014. The survey provides photometric data in six optical passbands. The only two SkyMapper bands we use in this work are the $i$ and $z$ bands, which have transmission curves that are similar to those of the eponymous SDSS filters. Differences in width and central wavelength are below 30~nm, and the SkyMapper passbands can be approximated by $(\lambda_{\rm cen}/{\rm fwhm})$ of (779/140) and (916/84) for $i$ and $z$, respectively.

When the first release of the SkyMapper Southern Survey \citep[DR1;][]{Wolf18a} became available in 2017, we started our search for high-redshift quasars, finding two of the objects reported here. When {\it Gaia} DR2 was released in April 2018, we accelerated our efforts, which led within a couple of days to the discovery of the quasar with the highest UV-optical luminosity known to date \citep{Wolf18b} and further objects. 

Recently, the second release (DR2) became available to Australian astronomers \citep{Onken19}. It mostly supersedes the first in sky coverage, depth and calibration quality, but also drops some of the older images by applying stricter quality limits. {\refbf While DR1 was complete to $i\approx 18.2$, DR2 goes to $i\approx 19.5$~mag on nearly 90\% of the Southern hemisphere. DR2 includes many more images, but its production was more selective in terms of image quality. As a result, 1\% of the sky area in DR1 has no $i$ or $z$ photometry in DR2. We note that two of the quasars we found before DR2 are in this lost area. One lacks photometry in the $z$ band, and the other is not in the DR2 catalogue at all. 

Although our work was started with DR1, we choose to report our results based on data from the superior DR2.
For the two objects without $z$ band photometry in DR2, we borrow their photometry from DR1 for the purpose of estimating their luminosity. Calibration differences between the two releases are small enough, such that if their $i$/$z$ magnitudes had been available in DR2, they would still have been selected as candidates. We expect the next release from SkyMapper to contain twice as many Main-Survey images and include those two objects again.} 

\subsection{Sample selection and quality flags}

The \texttt{master} table in the SkyMapper data release has cross-matches to {\it Gaia} and AllWISE. We consider SkyMapper objects that have a {\it Gaia} match within $2\arcsec$ (to accommodate centroid shifts when SkyMapper blends sources that are close together) as well as a {\it WISE} match within $4\arcsec$ distance. We also require photometric errors in {\it WISE} of ${\rm W1sigmpro}<0.1$ and ${\rm W2sigmpro}<0.2$, while we make no such requirements for {\it Gaia} and SkyMapper {\refbf since cuts in these bands are less critical when $R_p\la 18$}. We find that {\refbf $\sim$50}\% of {\it Gaia} sources at $|b|>20$~deg and $R_p<18$ have SkyMapper counterparts, driven by the survey footprint: SkyMapper covers nearly exactly the Southern hemisphere while {\it Gaia} is all-sky. Requiring a counterpart from the AllWISE all-sky table reduces the sample by $\sim 5$\%; the lost objects are most likely stars, which have bluer optical/mid-infrared colours than high-redshift quasars and thus the latter would be bright enough to be all retained by this rule. 

We exclude two regions centred on the two Magellanic Clouds with an area of $\sim 190$~deg$^2$. This leaves an effective search area of approx. {\refbf 12,500 deg$^2$, as} shown in Fig.~\ref{sky}. {\refbf Finally, we} remove 0.05\% of objects, whose SkyMapper DR2 flags indicate that their photometry is not reliable (${\rm flags}>3$). {\refbf SkyMapper flags up to value 255 come from Source Extractor \citep{BA96}, but we also use bespoke flags with higher values to indicate sources that are likely affected by either scattered light from nearby bright stars (with $i<6$~mag or $z<5$~mag), or possibly by uncorrected cosmic rays as indicated by higher light concentration than is allowed for a point source \citep[for details see][]{Onken19}.}

\begin{figure}
\begin{center}
\includegraphics[angle=0,width=\columnwidth,clip=true]{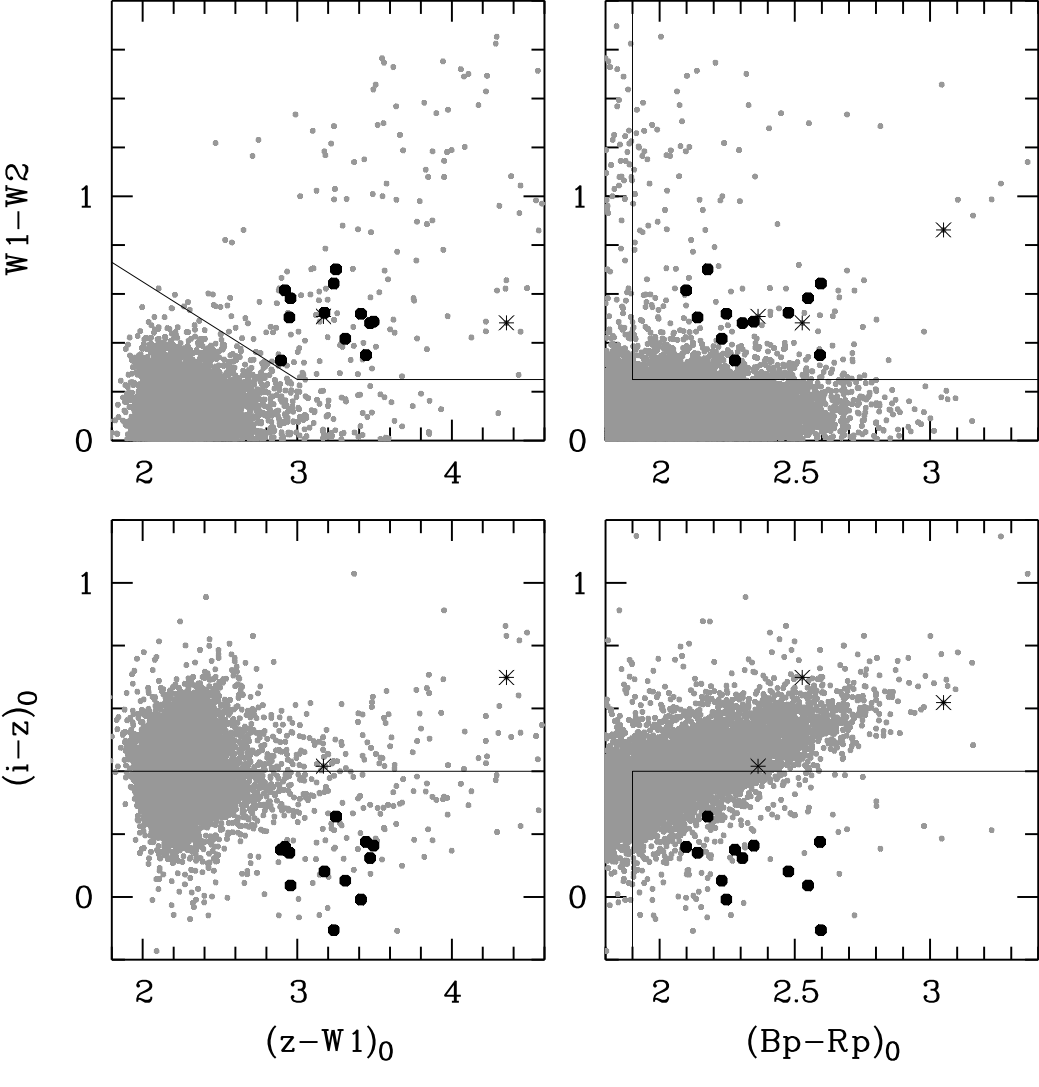} 
\caption{Colour-colour diagrams of objects with red {\it Gaia}/ {\it WISE} colours and magnitude $R_p=[15,18.2]$ after applying motion cuts. Small grey symbols are all objects, round black symbols are known $z\ga 4.5$ quasars and asterisks are known Carbon stars. Lines indicate the selection cuts.
\label{CCDs}}
\end{center}
\end{figure}

\begin{figure*}
\begin{center}
\includegraphics[angle=0,width=\textwidth,clip=true]{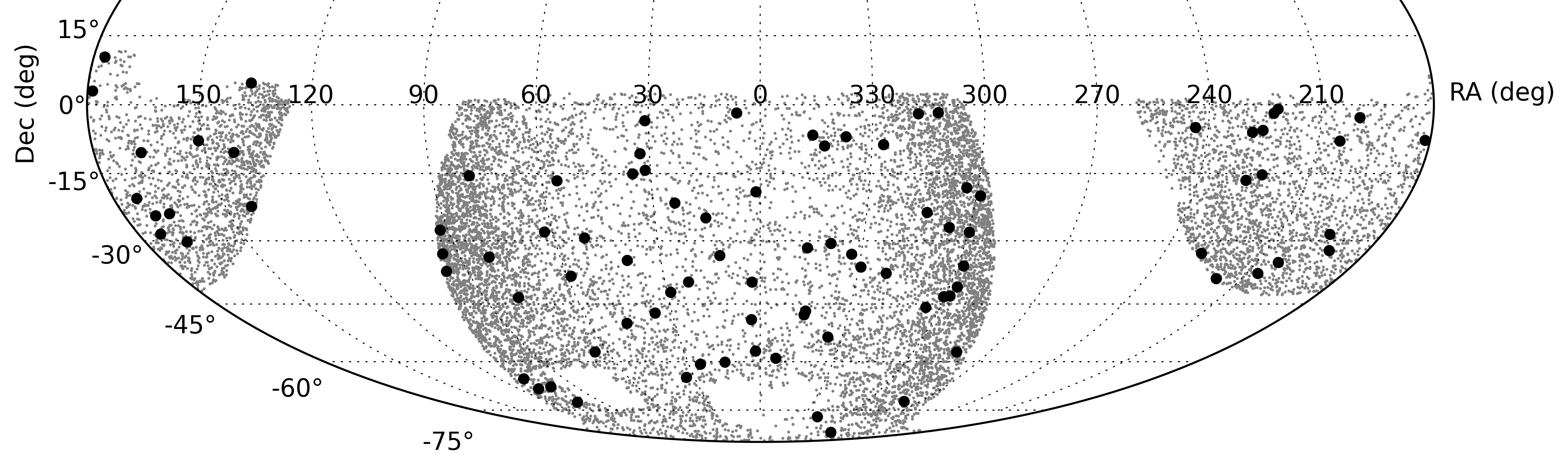} 
\caption{Sky map of objects with red {\it Gaia}/ {\it WISE} colours in the magnitude range $R_p=[15,18.2]$ after applying motion cuts. Small grey symbols are all {\refbf 13\,362} objects from $\sim ${\refbf 12\,500}~deg$^2$ of effective search area, and large black symbols show our {\refbf 92} final candidates selected by the cuts illustrated in Fig.~\ref{CCDs}.
\label{sky}}
\end{center}
\end{figure*}

\subsection{Colour selection}\label{coloursel}

We explore colour criteria for selecting high-redshift candidates using {\it Gaia} $B_p$ and $R_p$, SkyMapper $i$ and $z$, and {\it WISE} W1 and W2 bands. We start with a sample of objects that is expected to contain nearly all $z=[4.5,5.2]$ quasars with typical SEDs, using dereddened colour cuts of $(B_p-R_p)_0>1.8$, $(R_p-W1)_0>2$ and $W1-W2>0$, {\refbf as already motivated and chosen in Sect.~\ref{data_gaia}}. 

We corrected SkyMapper and {\it Gaia} magnitudes for interstellar extinction using the {\refbf reddening maps by \citet[hereafter SFD]{SFD98}}, while leaving {\it WISE} data unchanged given the long wavelength. We adopt reddening coefficients from \citet{Wolf18a}, whose $R_i=1.582$ and $R_z=1.2$ are used with $A=R \times E(B-V)_{\rm SFD}$, and for {\it Gaia} from \citet{CasaVDB18}, whose $R_{B_p}=3.374$ and $R_{R_p}=2.035$ are used with $A=R \times 0.86 E(B-V)_{\rm SFD}$. 

We search a magnitude range of $R_p=[15,18.2]$, given that the brightest known objects at $z>4$ have $R_p\approx 17$, and that we are interested in the brightest possible quasars in this first step of our work (in future work, we will reach deeper magnitudes as more telescope time for follow-up becomes available). Finally, we apply the stringent motion cuts from Section~\ref{data_gaia}. The resulting 13\,362 objects are shown in colour-colour diagrams in Fig.~\ref{CCDs}. 

The black round symbols mark twelve high-z quasars in our sample that were already known when we started this work. Four of these are at $z\approx 4.4$, just outside the notional redshift interval we are aiming for, but equally relevant. Three further objects with known spectra are Carbon stars, marked by asterisk symbols. 
{\refbf The latter} can have {\refbf extremely red} optical/mid-infrared colours, and {\refbf some} are known at $(z-W1)_0=[5,8]$, beyond our diagrams.

In the next step, we deselect the dense clump of Milky Way stars and define a candidate list for spectroscopic follow-up using further and tighter colour cuts. We aim for discovering our first batch of ultra-luminous high-redshift quasars and won't yet apply Bayesian selection approaches, but use conservative cuts aimed for purity; the spectroscopic follow-up of this candidate list is done with observations of one candidate at a time. In the near future, we will be able to follow up vastly larger candidate lists as ancillary science targets embedded in the Taipan Survey \citep{daCunha17}; then we will aim for a complete sample by penetrating the stellar locus. For now, however, we define candidates using the following criteria: 

\begin{eqnarray}
    (B_p-R_p)_0 > 1.9 ~ ~ {\rm ,} ~ ~ W1-W2>0.25 ~ ~ {\rm ,} ~ ~ (i-z)_0 < 0.4 \\
    ~ ~ {\rm and} ~ ~ W1-W2 > 0.25 + 0.4\times (3-(z-W1)_0)
\end{eqnarray}

These criteria selected 92 candidates with $R_p<18.2$ for spectroscopic follow-up. {\refbf None of the known $z>4.5$ quasars would be missed by these criteria, even if we pushed as faint as $R_p=18.8$, except for the object without $R_p$ data. Only at $R_p>18.8$ would we miss objects, presumably due to colour scatter from higher photometric errors.} Results are discussed in Tab.~\ref{tab_cands} and Sect.~\ref{results}. {\refbf Without the motion cuts another 3\,478 candidates would have been selected, so the {\it Gaia} data allowed us to compress the candidate list by a factor of $\sim$40 even in this part of colour space that is away from the main stellar locus.}

\begin{table*}
\centering
\caption{High-redshift quasars in our candidate sample. Four of the sources are radio-loud and detected in the radio survey NVSS; one of them (J052506.17-334305.6) is also detected in SUMSS.
{\refbf Errors on $R_p$, $z_{\rm PSF}$ and W1 range from 0.01 to 0.03 mag, well below the intrinsic variability of quasars. The average error on W1$-$W2 is 0.05 mag, but ranges from 0.04 to 0.09 mag. We only quote errors for the $H$~band, which are less homogenous.}
}
\label{tab_cands} 
\begin{tabular}{cccccccccl}
\hline \noalign{\smallskip}  
object  & redshift  	& $R_p$  & $z_{\rm PSF}$  & $H$  & W1 & W1$-$W2 & $M_{145}$ & $M_{300}$ & Comments \\
        &               & (Vega) & (AB)           & (Vega)           & (Vega) & (Vega)  & (AB)      & (AB) \\
\noalign{\smallskip} \hline \noalign{\smallskip}
\multicolumn{10}{c}{Quasars observed in this work} \\
\noalign{\smallskip} \hline \noalign{\smallskip}
 J000500.21-185715.3 &  4.56 &  18.050 &  18.33 & $ 16.72 \pm 0.06$ &  15.43 &  0.35 & $-27.85$ & $-28.14$ &   \\
 J000736.56-570151.8 &  4.25 &  16.989 &  17.21 & $ 15.73 \pm 0.03$ &  14.72 &  0.55 & $-28.88$ & $-29.03$ &   \\
 J001225.00-484829.8 &  4.59 &  17.541 &  17.59 & $ 16.01 \pm 0.02$ &  14.46 &  0.51 & $-28.33$ & $-28.86$ & $z_{\rm PSF}$ only in SMSS DR1 \\
 J013539.28-212628.2 &  4.94 &  17.886 &  17.73 & $ 15.84 \pm 0.15$ &  14.14 &  0.70 & $-28.59$ & $-29.27$ & NVSS $25.3\pm 0.9$~mJy \\
 J022306.73-470902.5 &  4.95 &  18.079 &  18.04 & $ 15.94 \pm 0.01$ &  14.55 &  0.59 & $-28.34$ & $-29.06$ &   \\
 J040914.87-275632.9 &  4.45 &  17.552 &  17.86 & $ 16.20 \pm 0.03$ &  14.68 &  0.49 & $-28.32$ & $-28.64$ & \citet{Schindler19b} \\
 {\refbf J051508.92-431853.7} &  4.61 &  18.064 &  18.06 & $ 16.25 \pm 0.04$ &  14.60 &  0.40 & $-28.03$ & $-28.64$ &   \\
 {\refbf J065522.86-721400.3} &  4.34 &  18.106 &  18.40 & $ 16.65 \pm 0.07$ &  14.87 &  0.39 & $-27.84$ & $-28.19$ &   \\
 J072011.68-675631.7 &  4.70 &  17.966 &  17.88 & $ 16.22 \pm 0.04$ &  14.29 &  0.53 & $-28.33$ & $-28.77$ &   \\
 J093032.57-221207.7 &  4.86 &  18.045 &  18.06 & $ 16.49 \pm 0.05$ &  15.43 &  0.55 & $-28.37$ & $-28.50$ &   \\
 J110848.48-102207.2 &  4.45 &  17.989 &  18.21 & $ 16.48 \pm 0.03$ &  15.16 &  0.50 & $-27.95$ & $-28.37$ & \citet{Schindler19b} \\
 J111054.69-301129.9 &  4.80 &  17.334 &  17.27 & $ 15.55 \pm 0.04$ &  14.27 &  0.52 & $-29.05$ & $-29.41$ & \citet{Yang19a} \\
 {\refbf J151204.27-055508.4} &  4.00 &  18.188 &  18.30 & $ 16.45 \pm 0.02$ &  14.91 &  0.38 & $-27.40$ & $-28.22$ &   \\
 {\refbf J151443.82-325024.8} &  4.81 &  17.758 &  17.83 & $ 15.82 \pm 0.03$ &  13.99 &  0.67 & $-28.96$ & $-29.34$ &   \\
 J211920.80-772252.9 &  4.52 &  17.399 &  17.56 & $ 16.03 \pm 0.04$ &  14.77 &  0.50 & $-28.58$ & $-28.87$ & only in SMSS DR1 \\
 J215728.21-360215.1 &  4.75 &  17.136 &  17.10 & $ 14.80 \pm 0.03$ &  13.10 &  0.51 & $-29.30$ & $-30.15$ & \citet{Wolf18b} \\
 J221111.55-330245.8 &  4.64 &  18.099 &  18.16 & $ 15.94 \pm 0.01$ &  13.85 &  0.61 & $-27.99$ & $-28.98$ &   \\
 J222759.64-065524.0 &  4.50 &  17.908 &  18.02 & $ 16.44 \pm 0.02$ &  14.76 &  0.42 & $-28.30$ & $-28.42$ &   \\
 J230349.19-063343.0 &  4.68 &  18.044 &  17.96 & $ 15.96 \pm 0.02$ &  14.55 &  0.47 & $-28.13$ & $-28.97$ &   \\
 J230429.88-313426.9 &  4.84 &  17.653 &  17.79 & $ 16.13 \pm 0.01$ &  14.62 &  0.61 & $-28.61$ & $-28.85$ &   \\
 J233505.85-590103.1 &  4.50 &  17.267 &  17.45 & $ 15.92 \pm 0.01$ &  14.43 &  0.45 & $-28.79$ & $-28.92$ &   \\
\noalign{\smallskip} \hline \noalign{\smallskip}
\multicolumn{10}{c}{Quasars known previously} \\
\noalign{\smallskip} \hline \noalign{\smallskip}
 J002526.83-014532.5 &  5.07 &  17.990 &  18.07 & $ 16.26 \pm 0.03$ &  14.80 &  0.64 & $-28.41$ & $-28.83$ &   \\
 J030722.86-494548.3 &  4.78 &  17.371 &  17.36 & $ 15.69 \pm 0.01$ &  14.38 &  0.58 & $-28.96$ & $-29.25$ &   \\
 J032444.28-291821.0 &  4.62 &  17.607 &  17.76 & $ 16.29 \pm 0.01$ &  14.84 &  0.33 & $-28.43$ & $-28.60$ & NVSS $236.5\pm 7.1$~mJy \\
 J035504.86-381142.5 &  4.58 &  17.232 &  17.42 & $ 15.59 \pm 0.10$ &  13.94 &  0.48 & $-28.75$ & $-29.27$ &   \\
 J041950.93-571612.9 &  4.46 &  18.132 &  18.44 & $ 16.67 \pm 0.04$ &  15.01 &  0.52 & $-27.80$ & $-28.16$ &   \\
 J052506.17-334305.6 &  4.42 &  18.011 &  18.18 & $ 16.31 \pm 0.04$ &  14.90 &  0.70 & $-27.80$ & $-28.51$ & NVSS $188.3\pm 5.7$~mJy \\
 J071431.39-645510.6 &  4.44 &  17.741 &  17.91 & $ 16.33 \pm 0.04$ &  14.85 &  0.50 & $-28.27$ & $-28.53$ & gravitationally lensed \\
 J090527.39+044342.3 &  4.40 &  17.812 &  18.24 & $ 16.27 \pm 0.03$ &  14.70 &  0.49 & $-27.84$ & $-28.55$ & blended in SMSS DR2 table \\
 J120523.14-074232.6 &  4.69 &  17.853 &  17.92 & $ 16.22 \pm 0.03$ &  14.70 &  0.52 & $-28.37$ & $-28.70$ &   \\
 J130031.13-282931.0 &  4.71 &  18.022 &  18.05 & $ 16.35 \pm 0.04$ &  14.51 &  0.35 & $-28.23$ & $-28.62$ &   \\
 J145147.04-151220.1 &  4.76 &  17.016 &  17.10 & $ 15.35 \pm 0.01$ &  13.69 &  0.42 & $-29.29$ & $-29.63$ & NVSS $28.5\pm 1.0$~mJy \\
 J211105.60-015604.1 &  4.85 &  18.007 &  17.96 & $ 16.39 \pm 0.03$ &  15.02 &  0.61 & $-28.48$ & $-28.61$ &   \\
 J214725.70-083834.5 &  4.59 &  17.795 &  17.97 & $ 16.38 \pm 0.04$ &  15.00 &  0.61 & $-28.22$ & $-28.51$ &   \\
{\refbf J223953.66-055220.0} &  4.56 &         &  17.71 & $ 15.90 \pm 0.02$ &  14.09 &  0.47 & $-28.46$ & $-28.98$ & no Gaia $R_p$ measurement \\
\noalign{\smallskip} \hline
\end{tabular}
\end{table*}


\subsection{Spectroscopy at the ANU 2.3m-telescope}

We took spectra of the candidates with the Wide Field Spectrograph \citep[WiFeS;][]{Dopita10} on the ANU 2.3m-telescope at Siding Spring Observatory. We used {\refbf 18} nights, most of which were useful (2018 April 29, Sep 16 to 18, Dec 4, 5; 2019 March 11 to 13, 30, April 1, June 30, July 1 to 5, {\refbf Aug 29}). Two quasars were identified already with earlier observations on 2017 December 21 during a pilot search that had not yet benefited from {\it Gaia} data. 

We used the WiFeS B3000 and R3000 gratings in the blue and red arm, which between them cover the wavelength range from 3600~\AA \ to 9800~\AA \ at a resolution of $R=3000$. Exposure times ranged from 600~sec to 1\,800~sec, and observing conditions varied significantly in terms of cloud cover and seeing. 

The data were reduced using the Python-based pipeline PyWiFeS \citep{Childress14}. PyWiFeS calibrates the raw data with bias, arc, wire, internal-flat and sky-flat frames, and performs flux calibration and telluric correction with standard star spectra. Flux densities were calibrated using several standard stars throughout the year, which are usually observed on the same night. We then extract spectra from the calibrated 3D cube using a bespoke algorithm for fitting and subtracting the sky background after masking sources in a white-light stack of the 3D cube. 

{\refbf We obtained spectra with for all candidates except the objects reported in the literature. Reduced spectra were visualised with the MARZ software \citep{Hinton16} and quasar templates were overplotted to aid the redshift determination. We estimate redshifts from broad Si~{\sc iv} and C~{\sc iv} lines where available, and consider the blue edge of the Ly~$\alpha$ line when necessary. We estimate that redshift uncertainties range from 0.01 to $\sim $0.05 for the weak-lined objects. Despite good signal, four of the 92 candidates from our sample remain currently unclassified. Their spectra are unlike anything we have seen before, and we deem them unlikely to be high-redshift quasars}.

\begin{figure*}
\begin{center}
\includegraphics[angle=270,width=0.32\textwidth,clip=true]{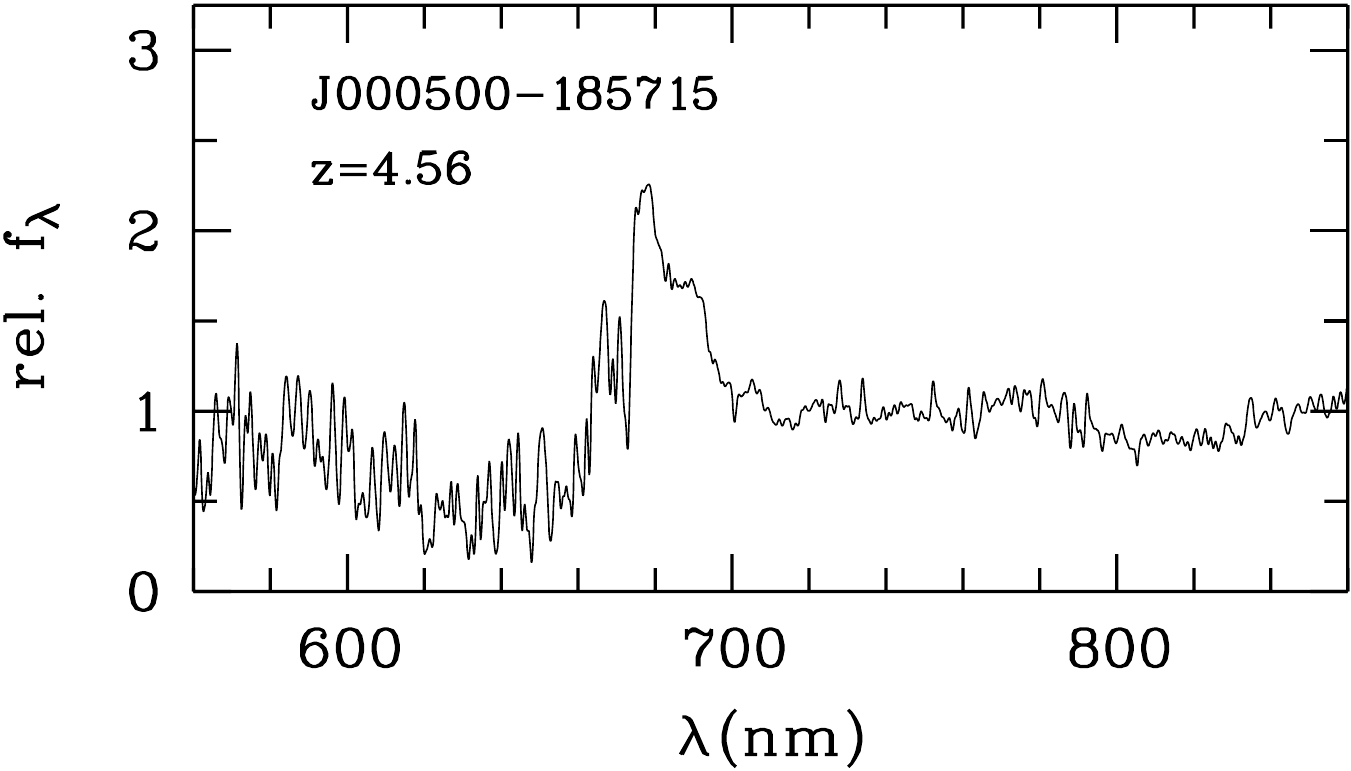}
\includegraphics[angle=270,width=0.32\textwidth,clip=true]{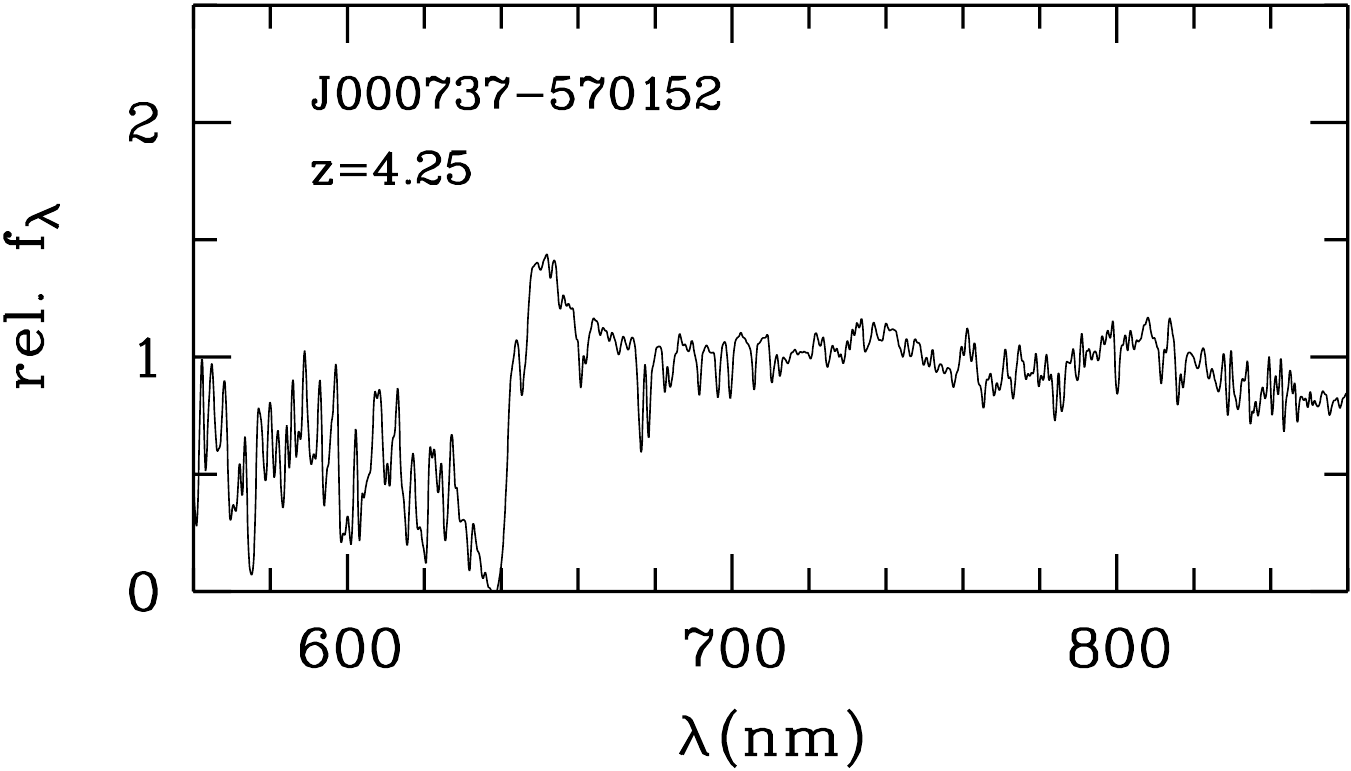}
\includegraphics[angle=270,width=0.32\textwidth,clip=true]{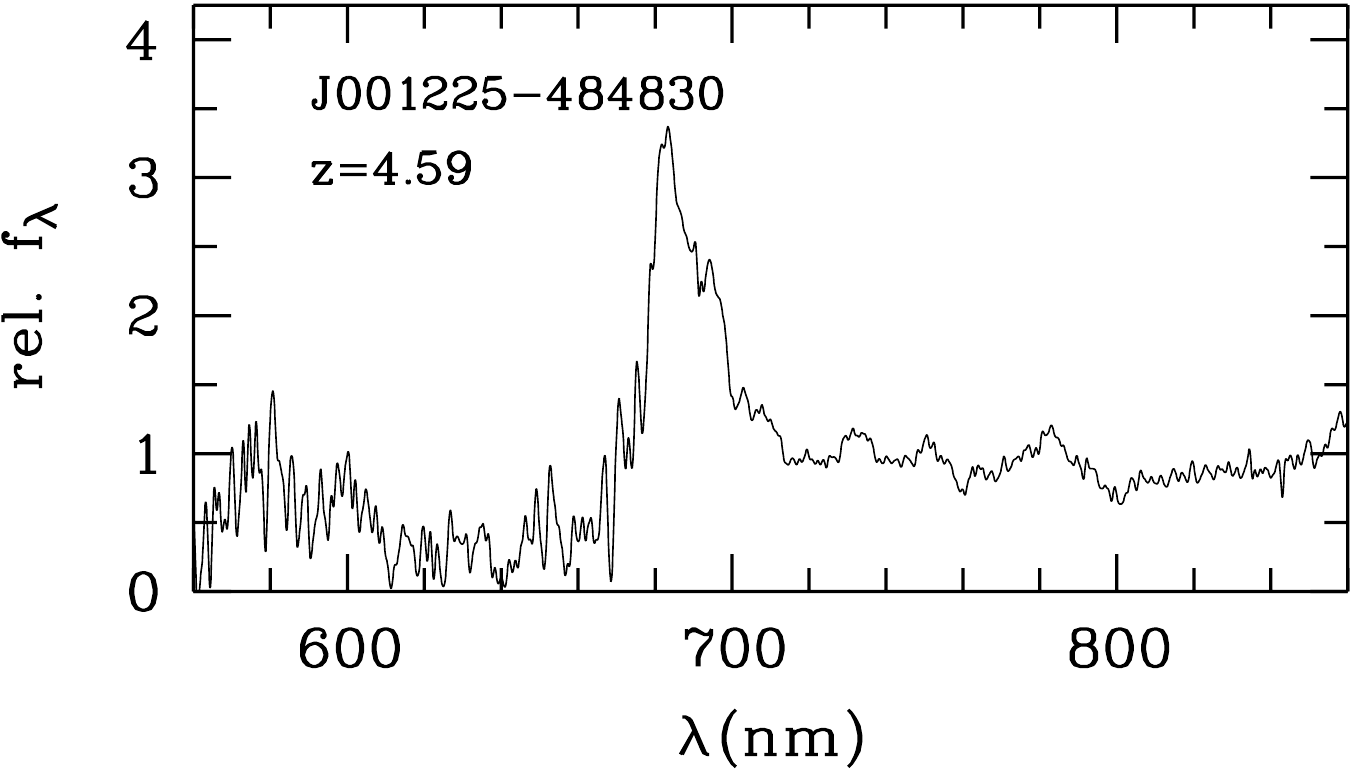}
\includegraphics[angle=270,width=0.32\textwidth,clip=true]{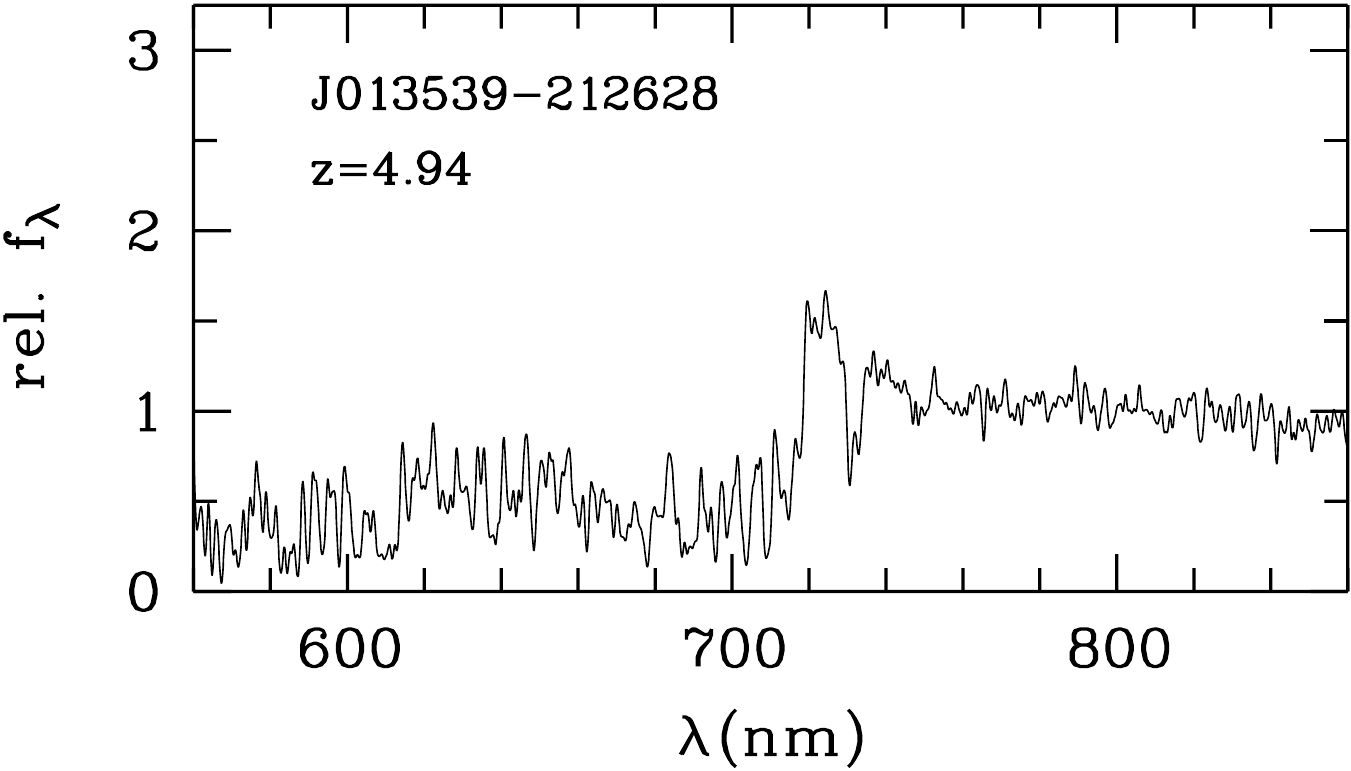}
\includegraphics[angle=270,width=0.32\textwidth,clip=true]{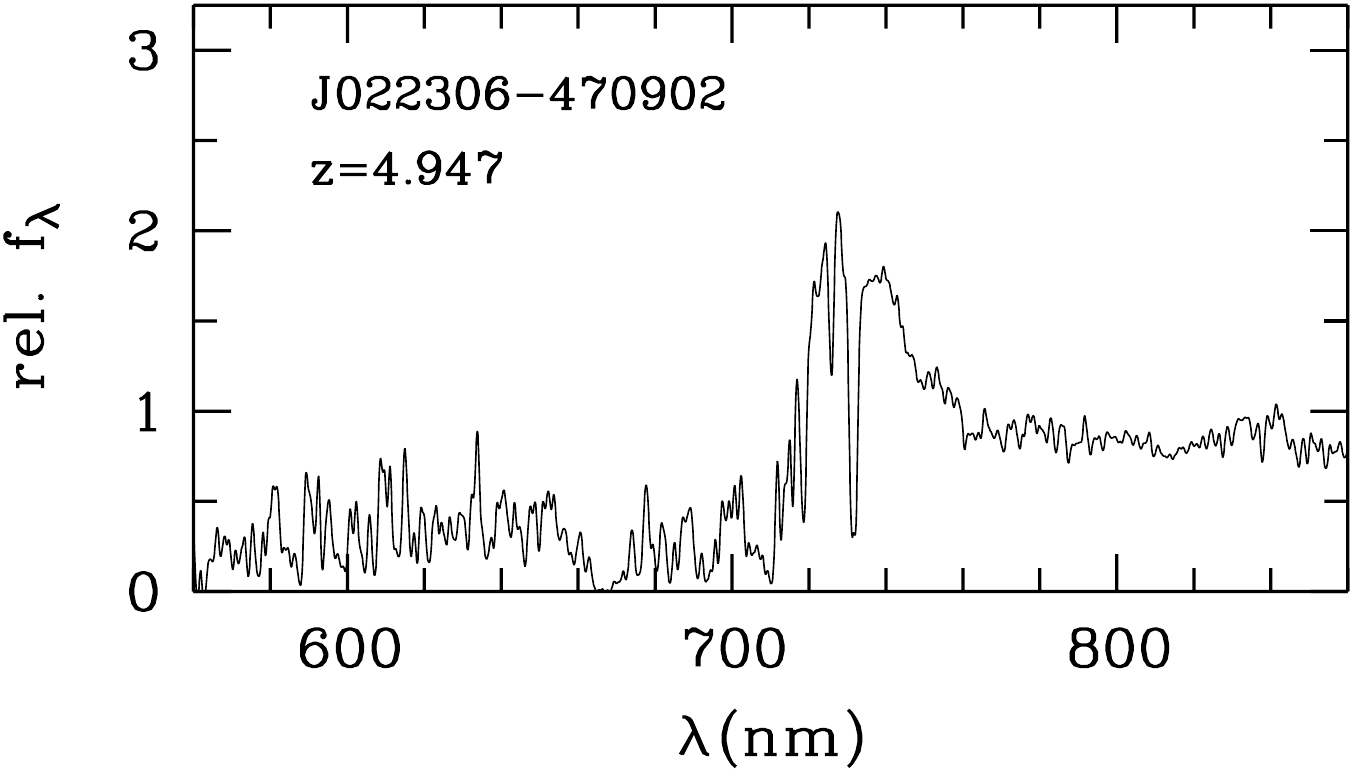}
\includegraphics[angle=270,width=0.32\textwidth,clip=true]{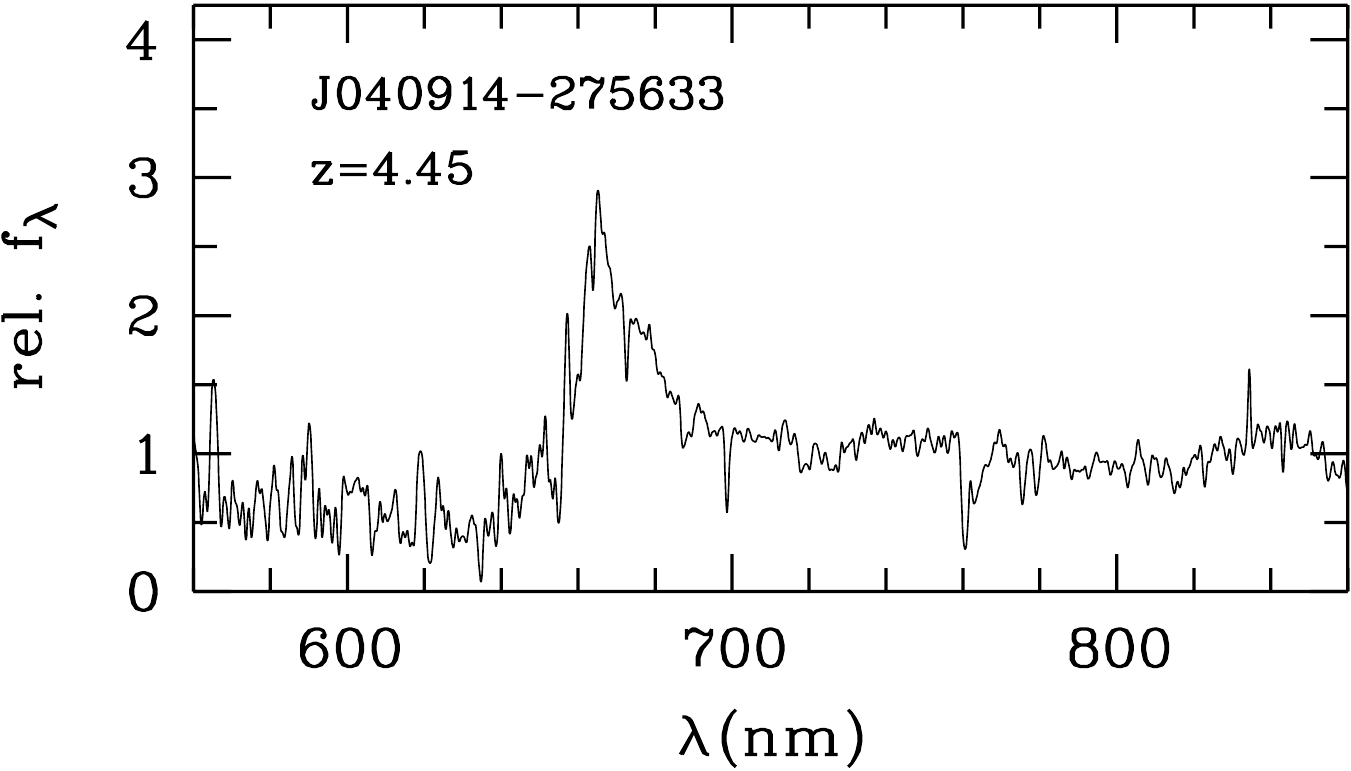}
\includegraphics[angle=270,width=0.32\textwidth,clip=true]{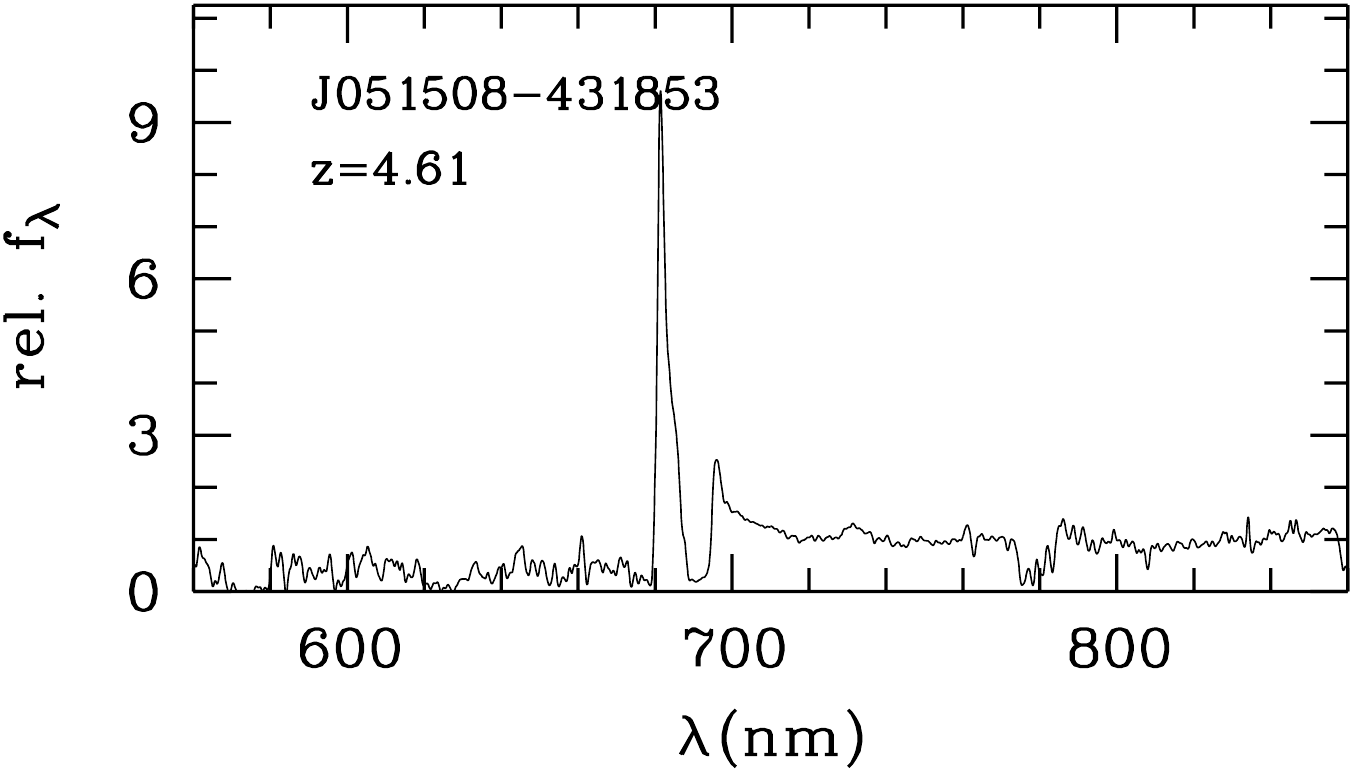}
\includegraphics[angle=270,width=0.32\textwidth,clip=true]{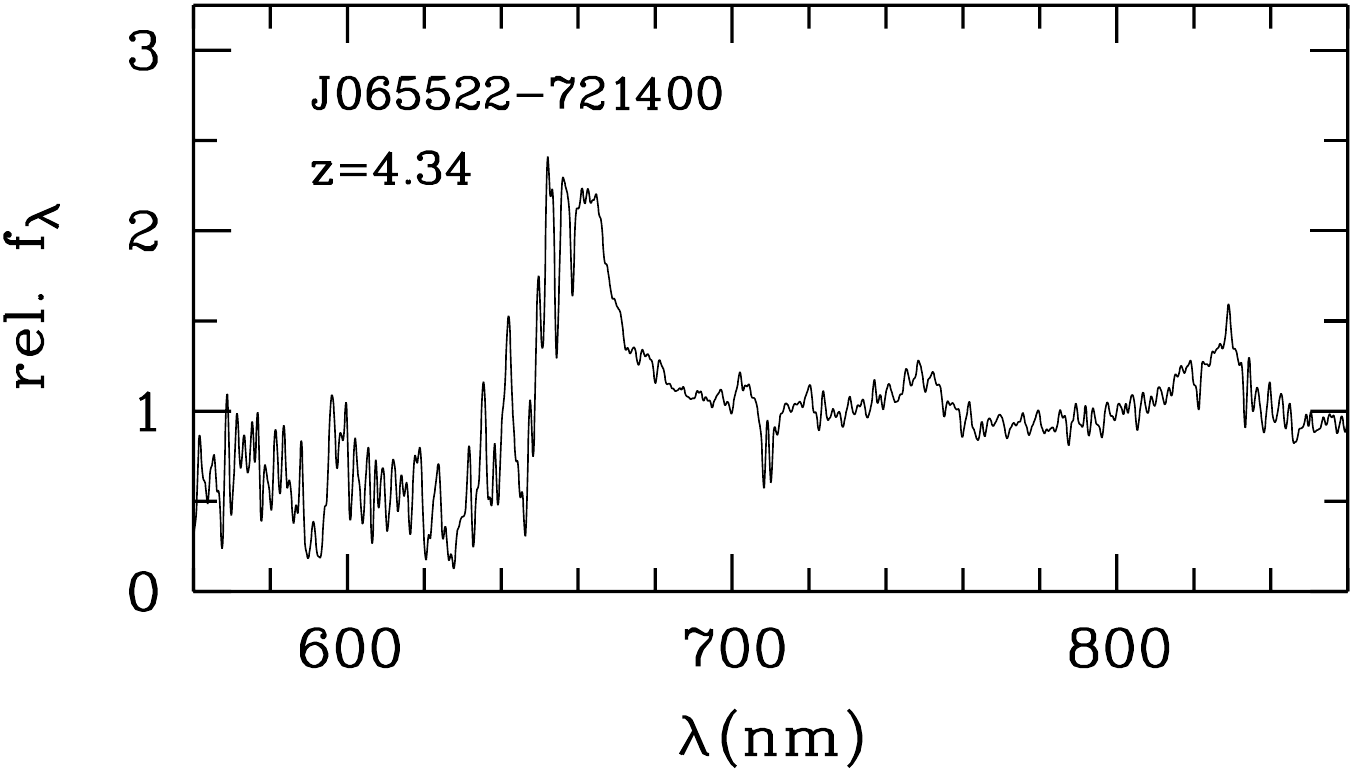}
\includegraphics[angle=270,width=0.32\textwidth,clip=true]{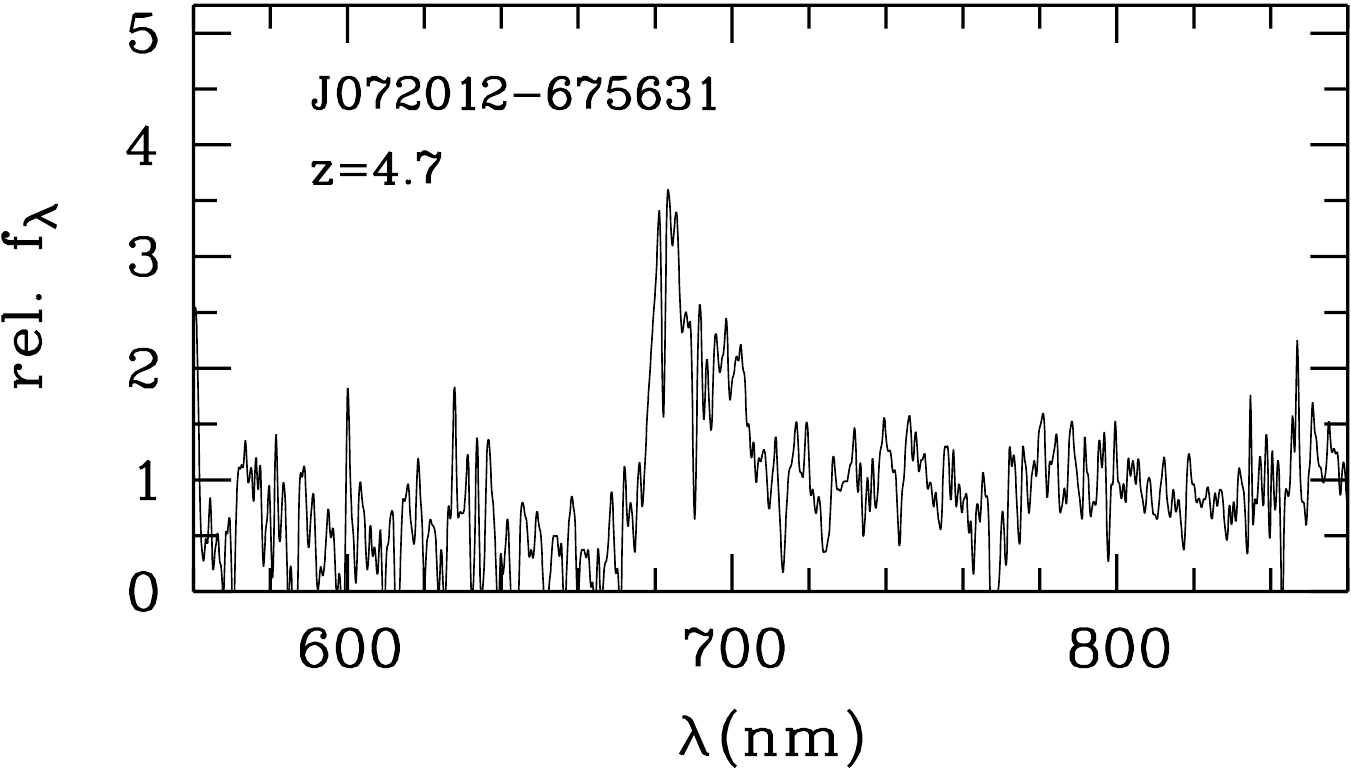}
\includegraphics[angle=270,width=0.32\textwidth,clip=true]{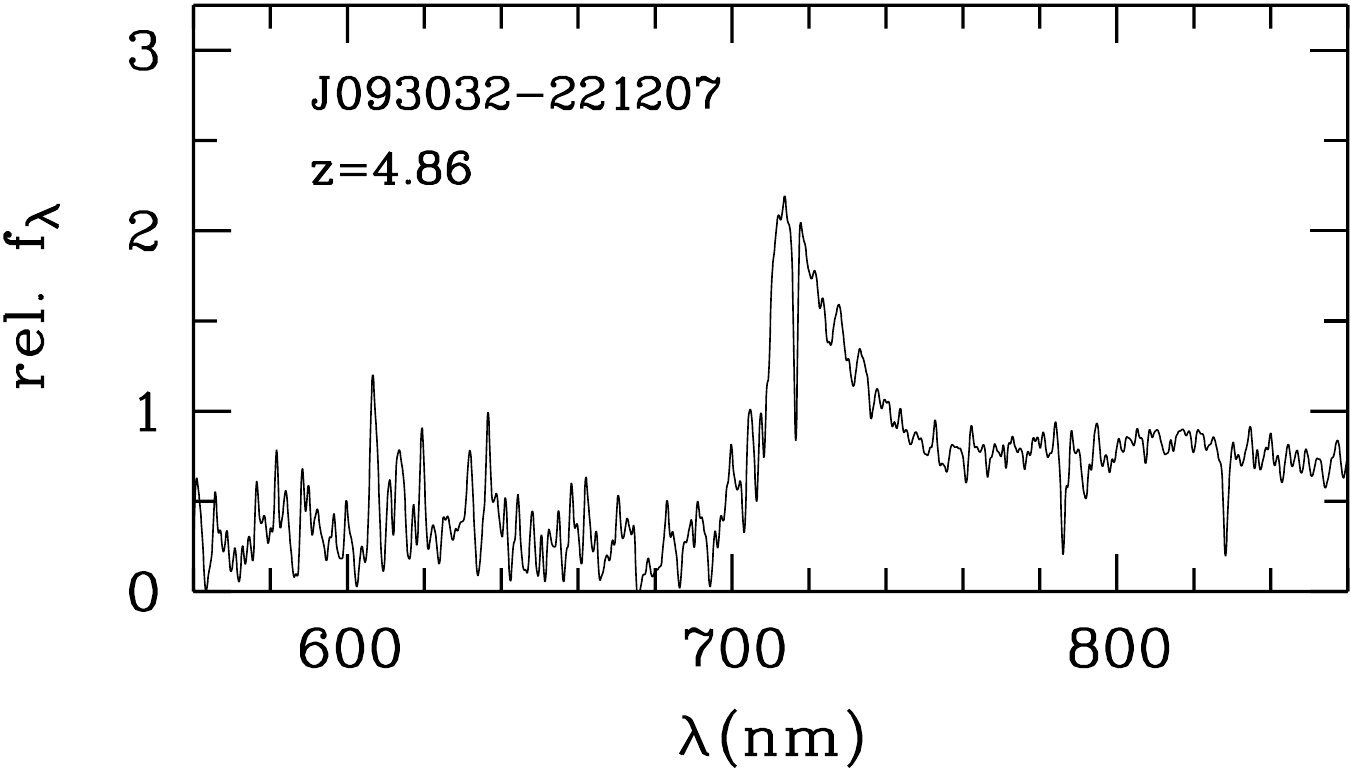}  
\includegraphics[angle=270,width=0.32\textwidth,clip=true]{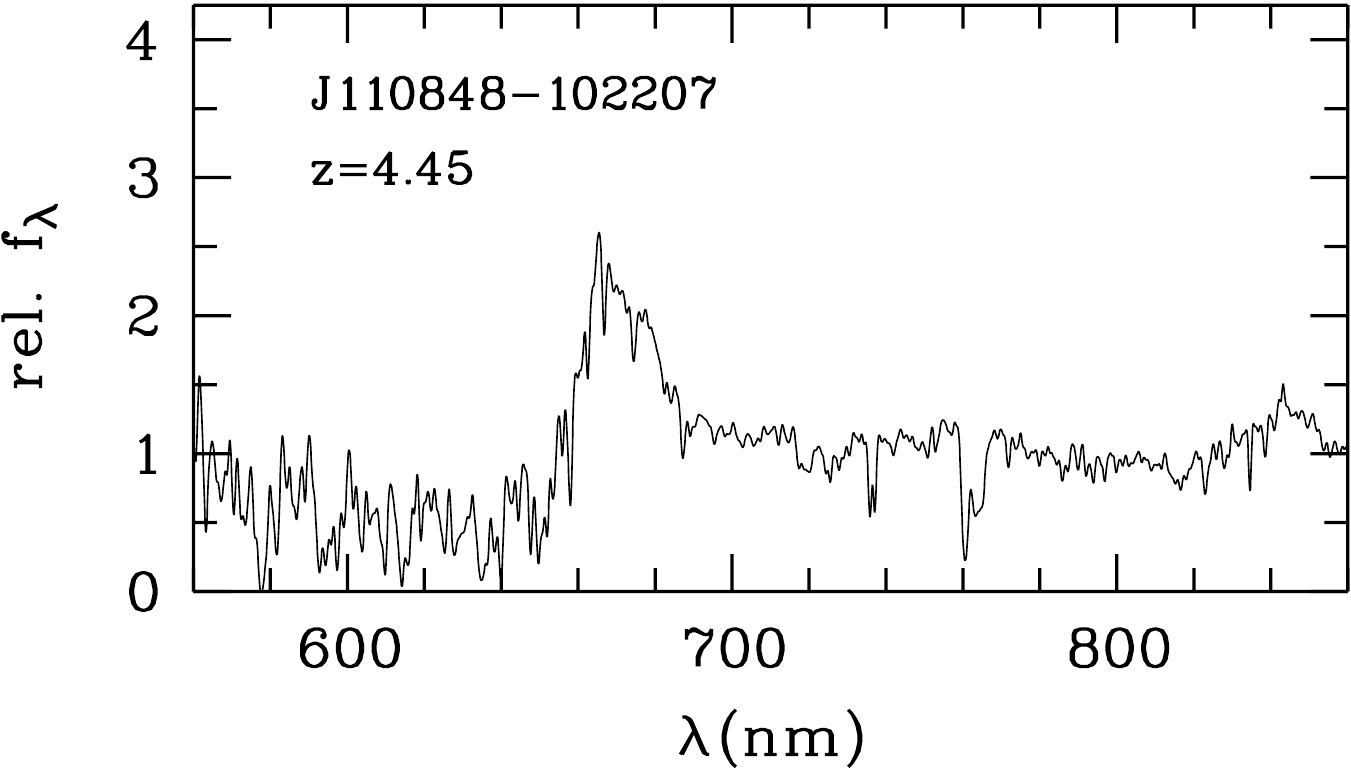}
\includegraphics[angle=270,width=0.32\textwidth,clip=true]{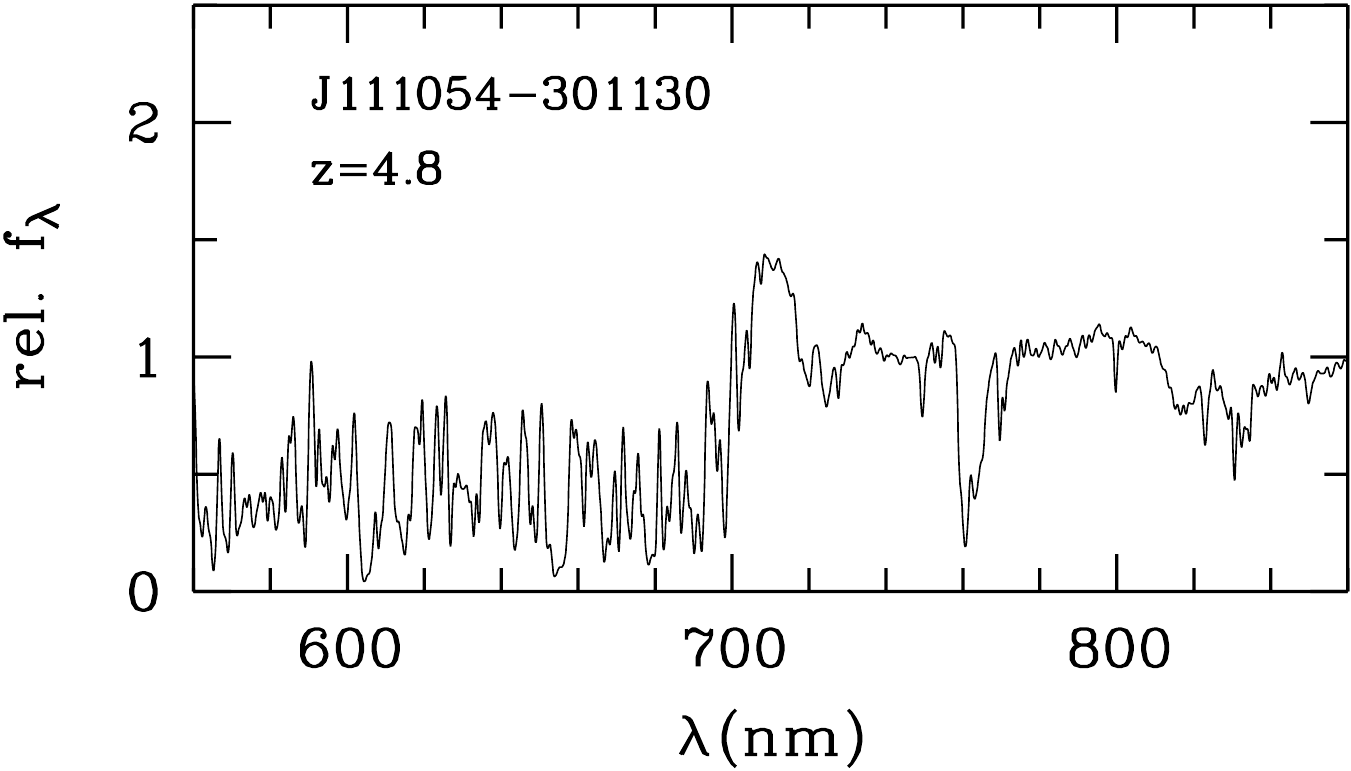}
\includegraphics[angle=270,width=0.32\textwidth,clip=true]{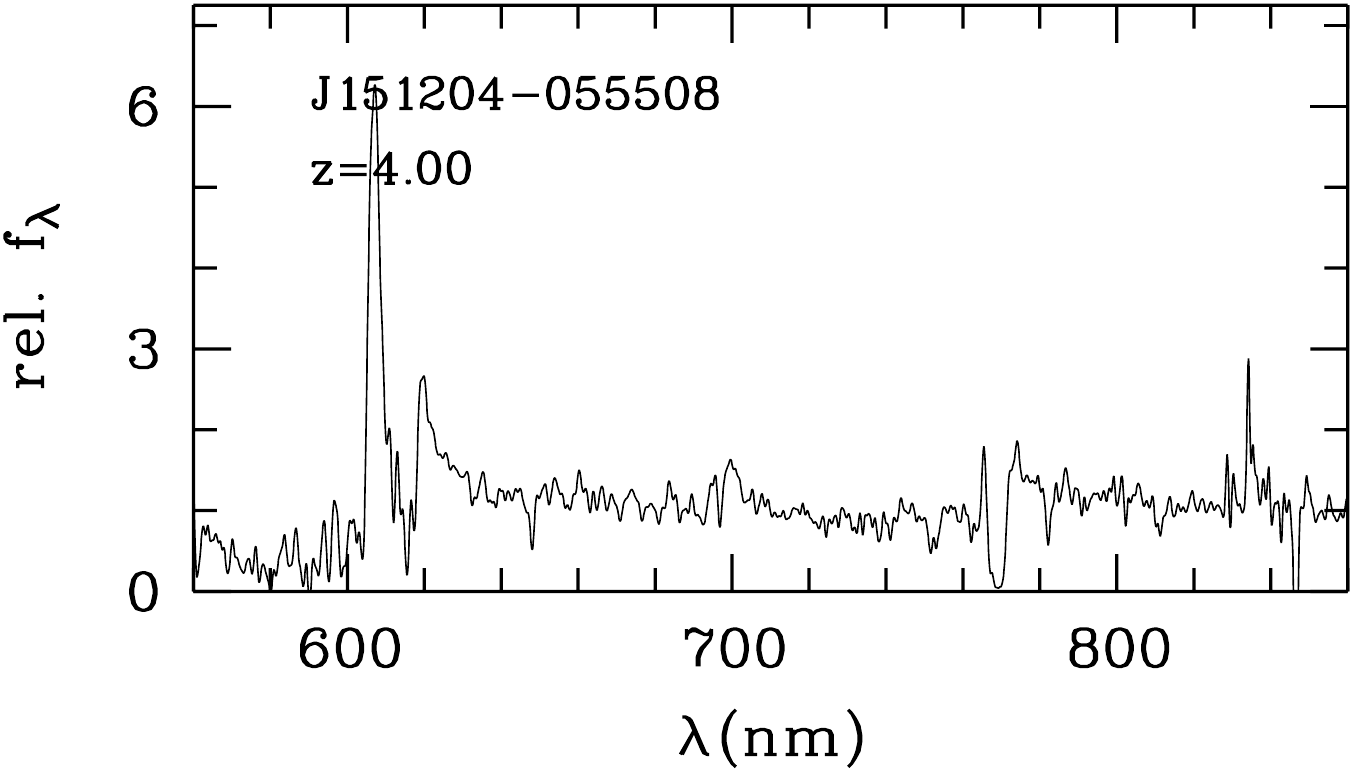}
\includegraphics[angle=270,width=0.32\textwidth,clip=true]{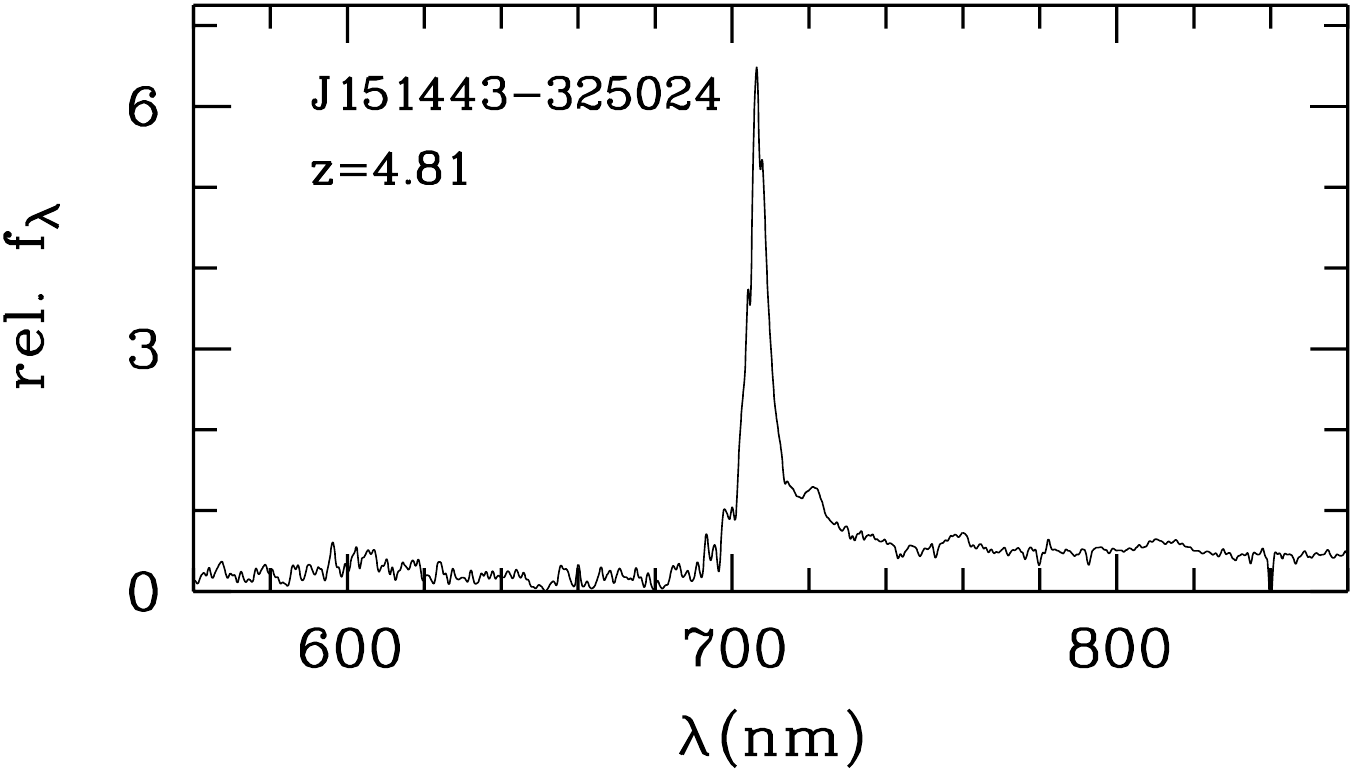}
\includegraphics[angle=270,width=0.32\textwidth,clip=true]{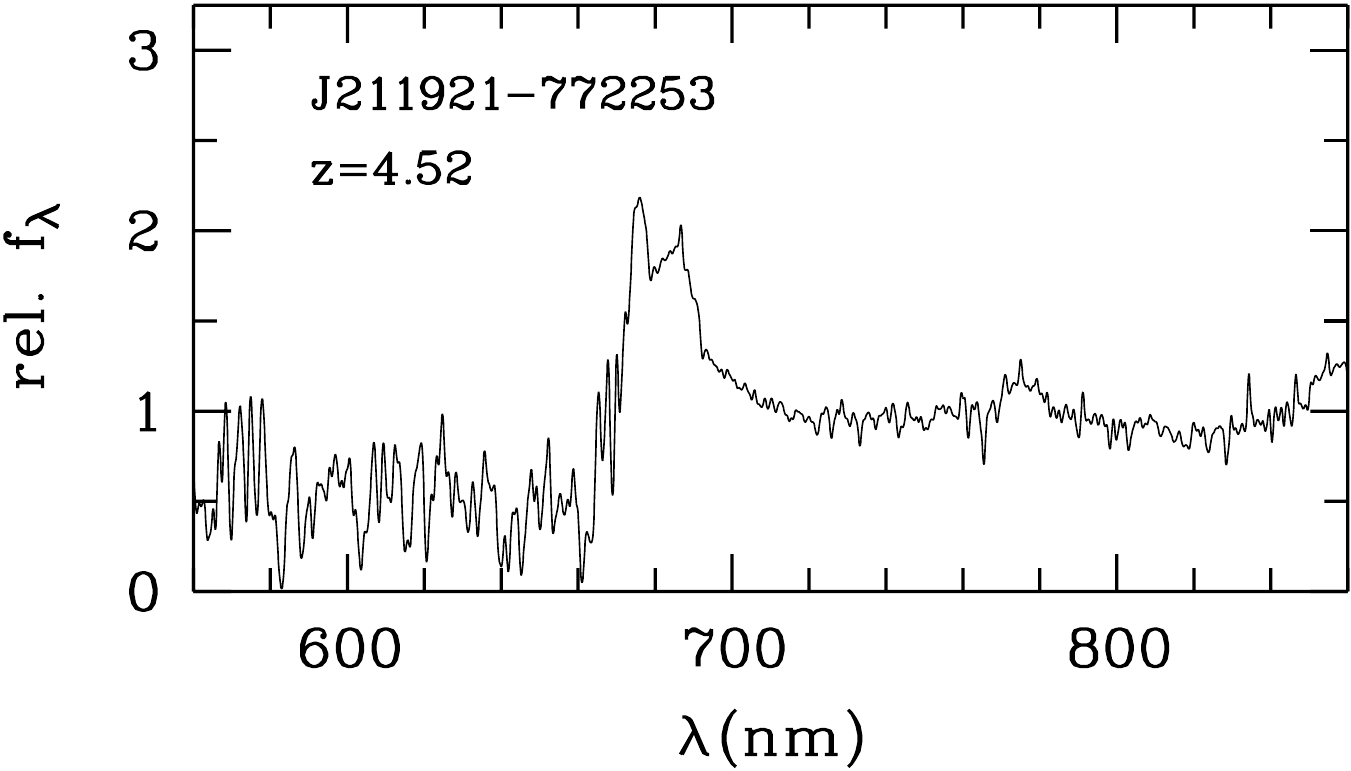}
\includegraphics[angle=270,width=0.32\textwidth,clip=true]{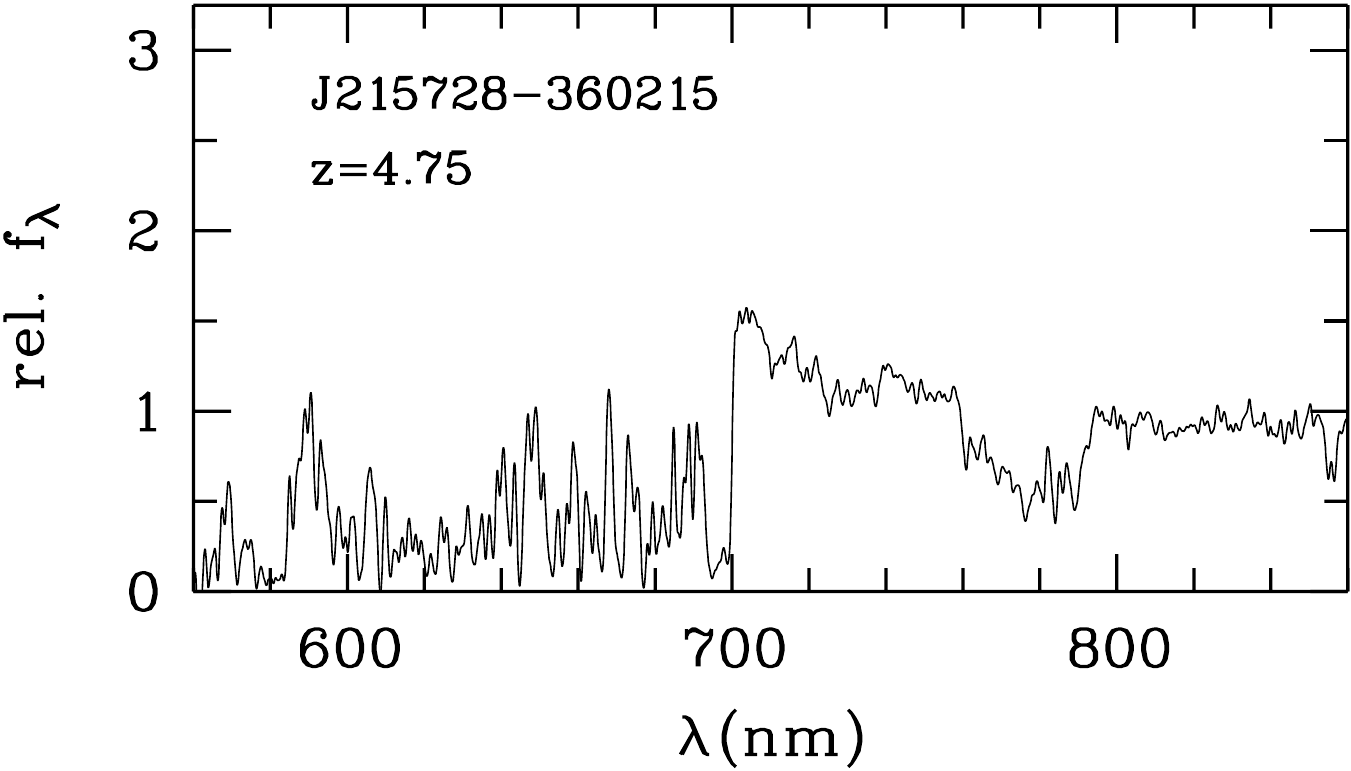}
\includegraphics[angle=270,width=0.32\textwidth,clip=true]{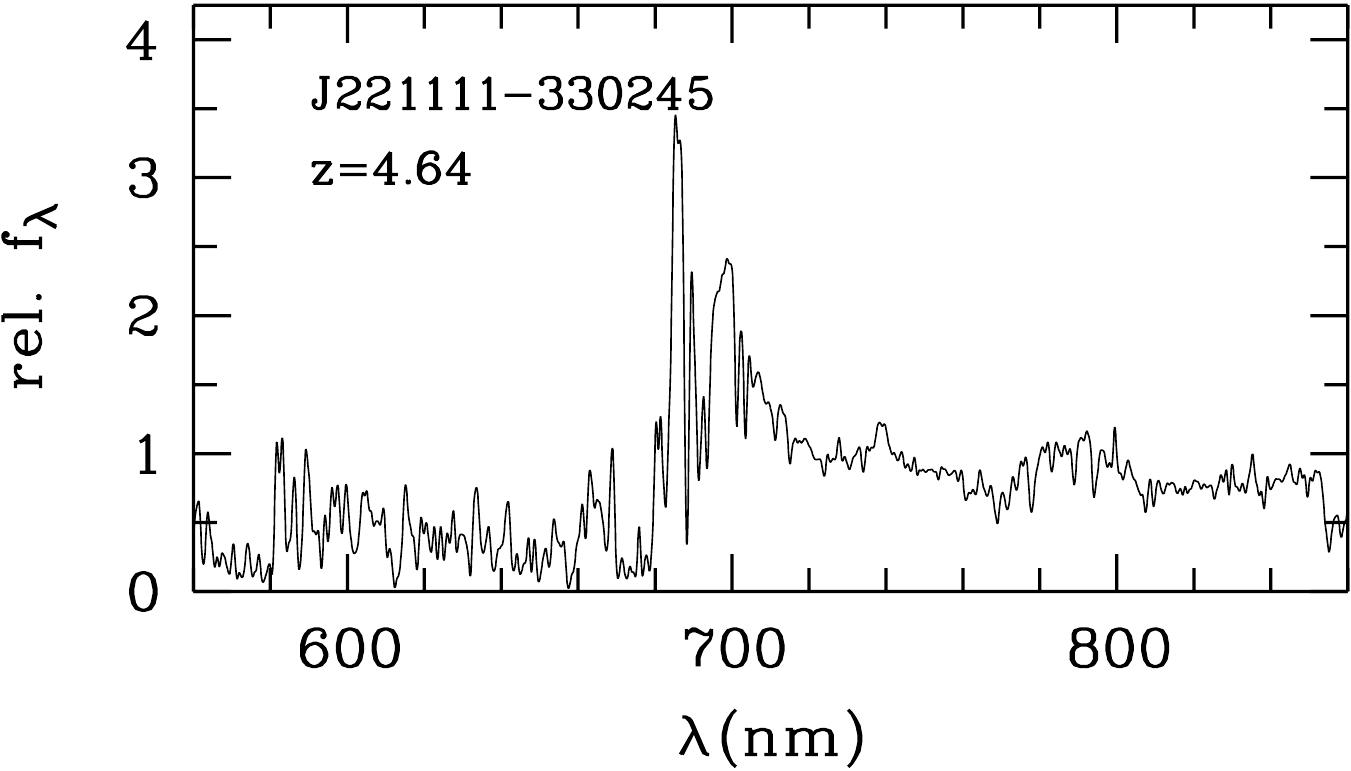}  \includegraphics[angle=270,width=0.32\textwidth,clip=true]{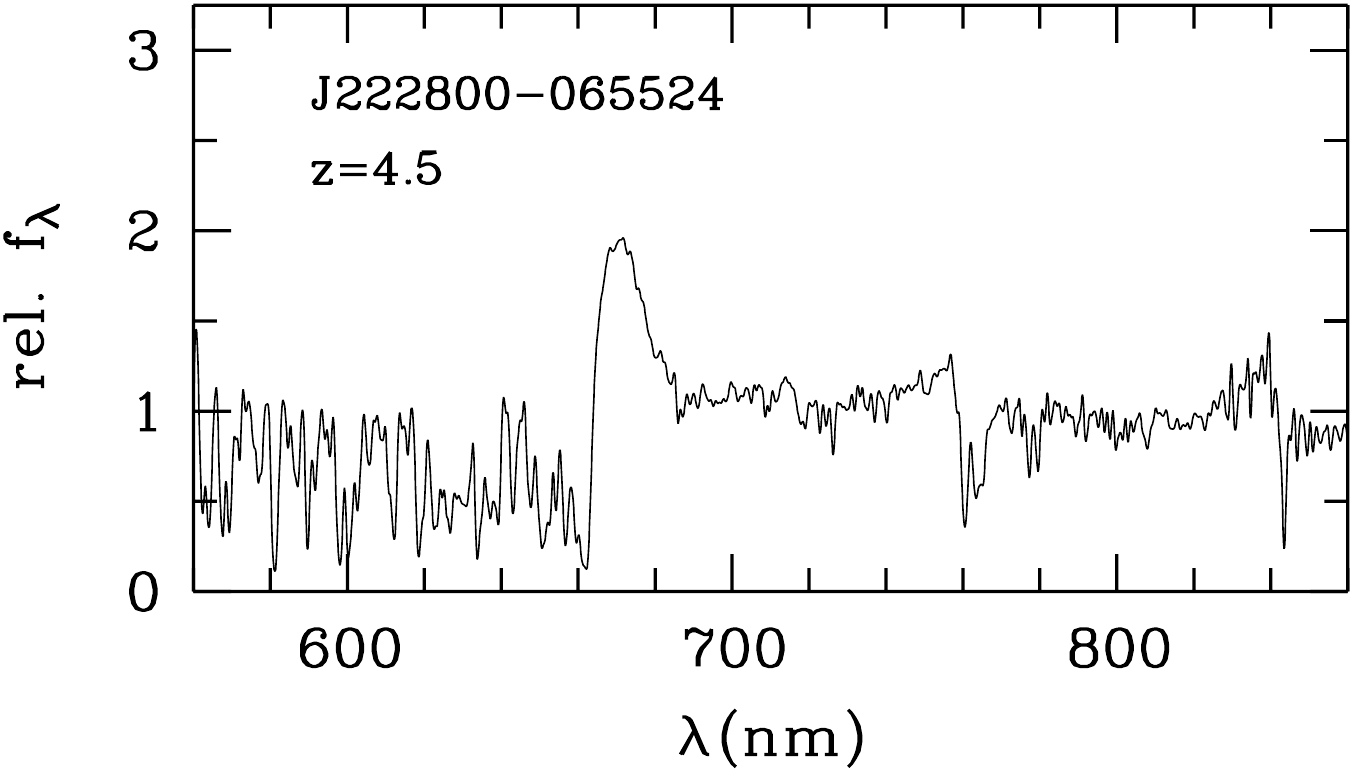}
\includegraphics[angle=270,width=0.32\textwidth,clip=true]{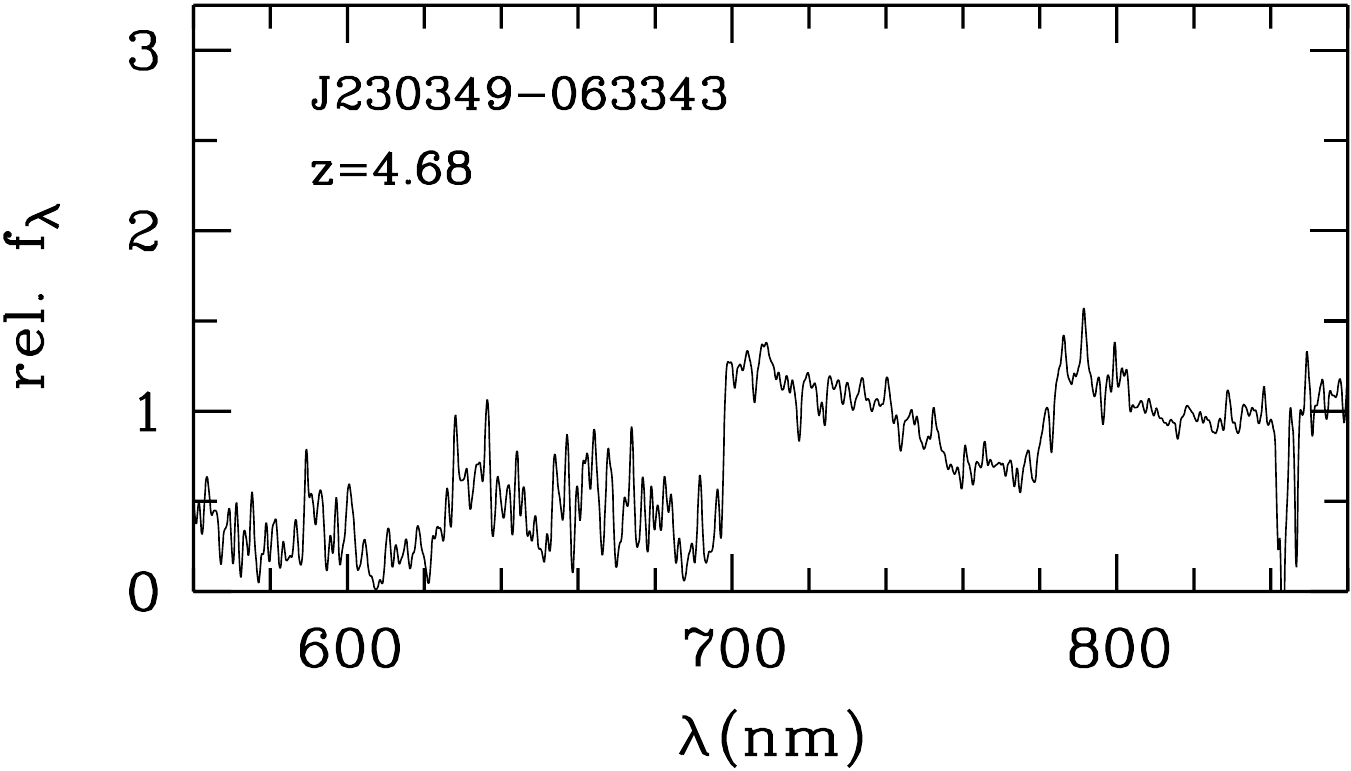}  \includegraphics[angle=270,width=0.32\textwidth,clip=true]{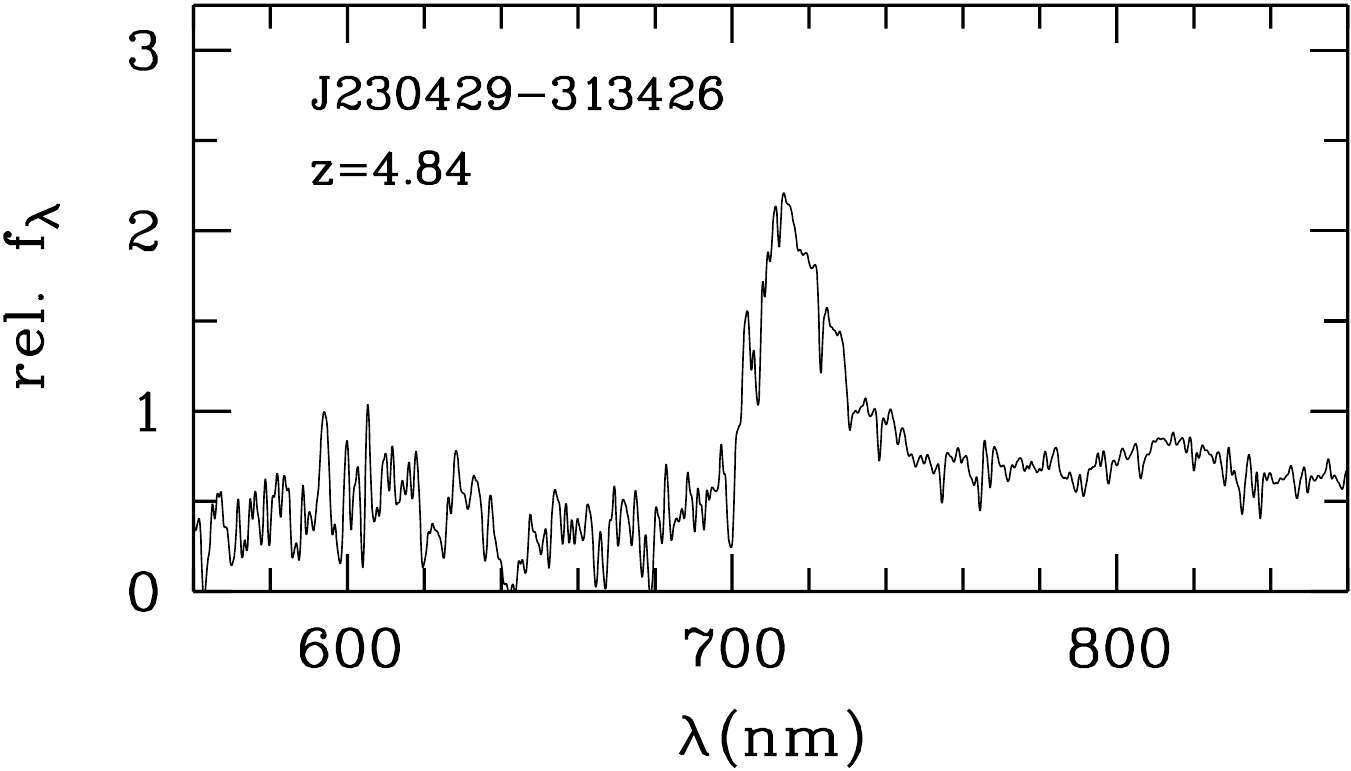}  \includegraphics[angle=270,width=0.32\textwidth,clip=true]{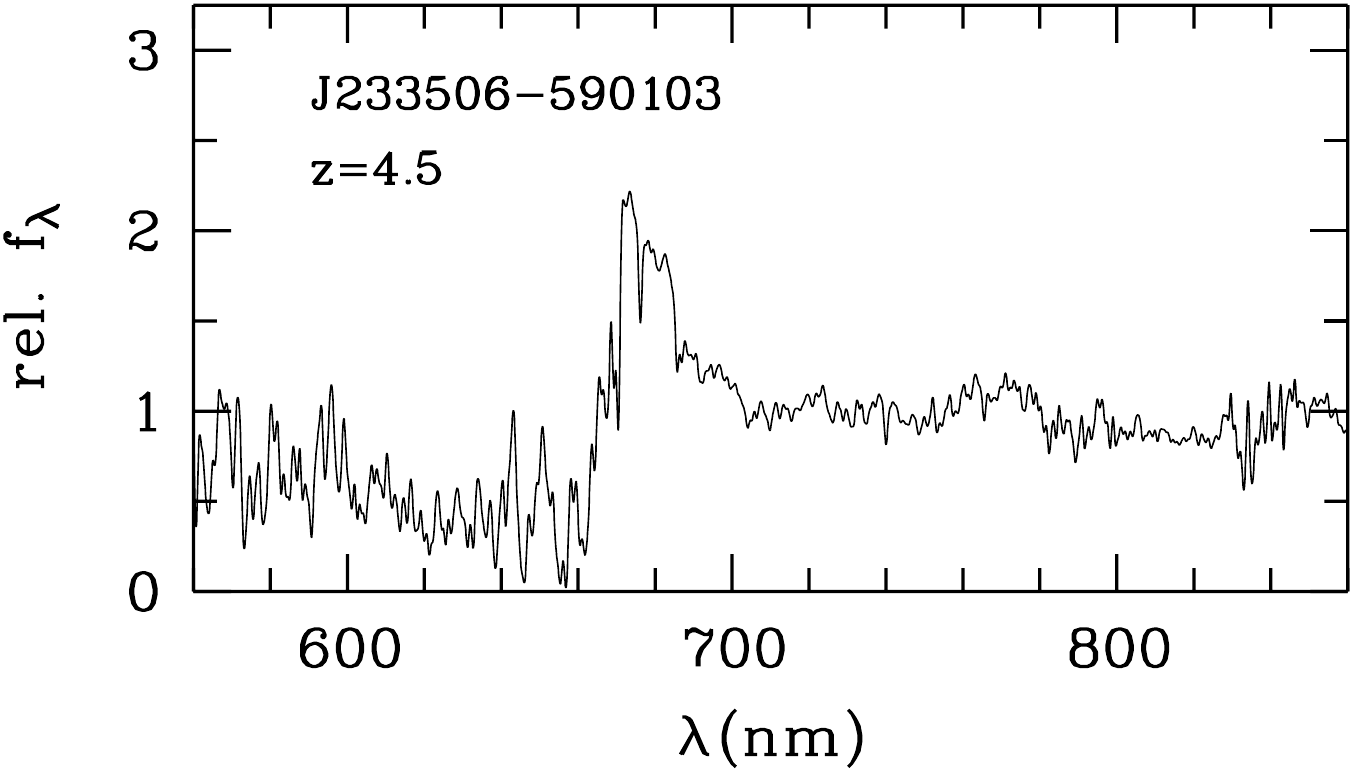}
\caption{Gallery of {\refbf 21 quasar spectra with $z\ge 4$} observed {\refbf by us. The two quasars with weak emission lines have clearly visible Lyman forests} in the 2D spectra. {\refbf Example spectra from the 24 red or unusual quasars at lower redshift are shown in Fig.~\ref{gallery_lozquasars}.}
\label{gallery_hizquasars}}
\end{center}
\end{figure*}

\begin{figure}
\begin{center}
\includegraphics[angle=270,width=\columnwidth,clip=true]{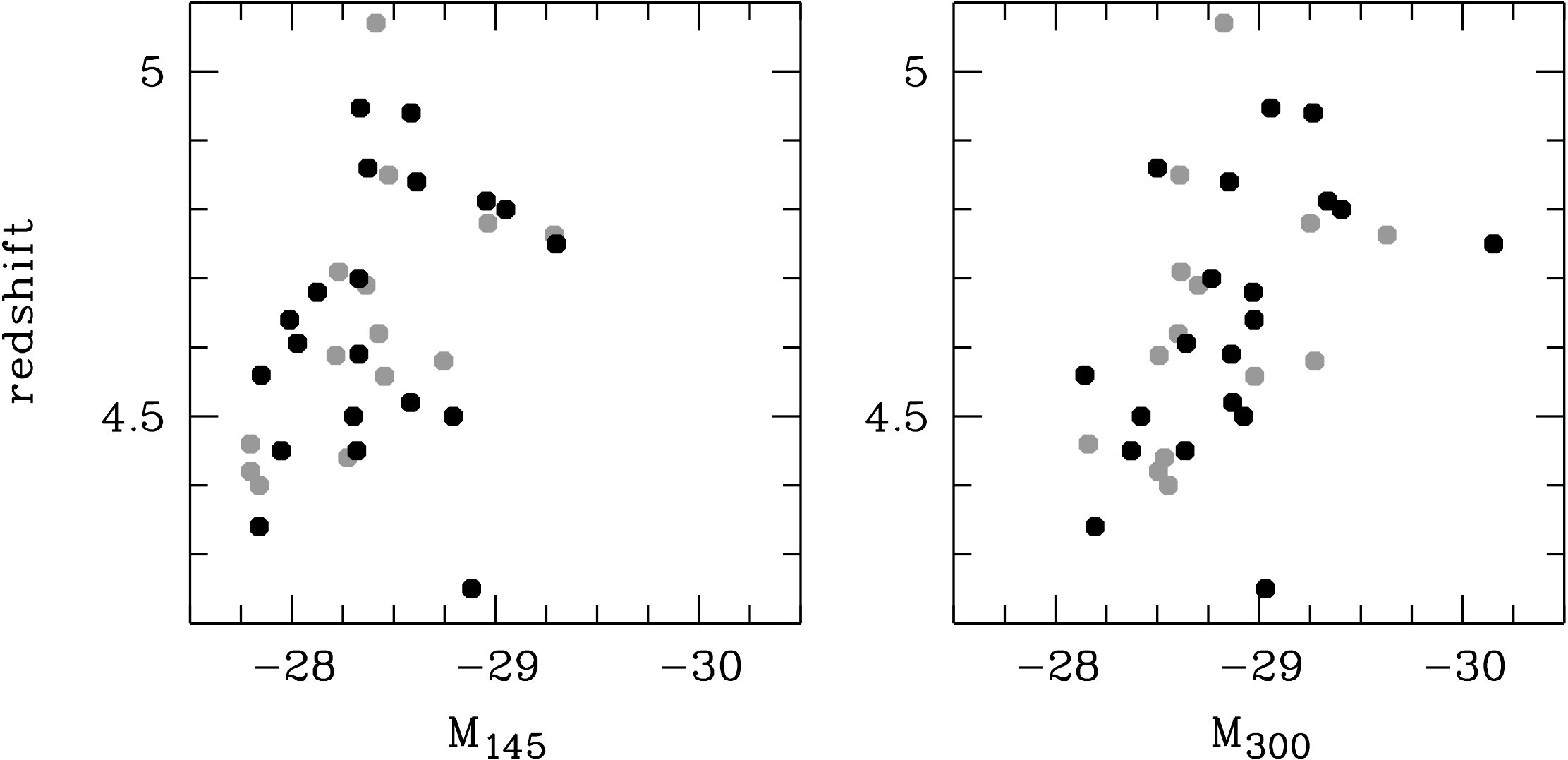} 
\caption{Ultra-luminous high-redshift quasars {\refbf with $R_p<18.2$}, known before this work (grey) and identified in this work (black, {\refbf see Table~\ref{tab_cands})}. Our sample improves especially in the regime of high bolometric luminosity, for which $M_{300}$ is a useful proxy.
\label{Lz}}
\end{center}
\end{figure}

\section{Results}\label{results}

\subsection{High-redshift quasars among the candidates}

The spectroscopy revealed {\refbf 21} ultra-luminous quasars at $z\ge 4$ (see Fig.~\ref{gallery_hizquasars}), of which {\refbf 18} are newly identified. The most luminous object among the new {\refbf 18} was already reported in \citet{Wolf18b}. Three more objects in our sample were independently observed by other teams and published in the literature only after our observations were done: \citet{Yang19a} revealed a $z=4.8$ quasar and \citet{Schindler19b} found two more quasars at $z=4.45$. {\refbf Three} of the newly identified objects are at slightly lower {\refbf redshifts of $z=4$, $z=4.25$ and $z=4.34$}, while the others are at $z\ge 4.5$ as expected. 

In the bottom part of Table 1 we list {\refbf 13} further high-redshift quasars in our sample, {\refbf and one outside our sample}, that were known already when our observations commenced: {\refbf ten} of them are in the target redshift window of $z=[4.5,5.2]$ and four more at slightly lower redshifts. Altogether then, {\refbf 34} objects out of {\refbf 92} candidates with good spectra are quasars at high redshifts of $z\ge 4$. 

As a result of this work, we have {\refbf lifted} the number of known {\refbf bright ($R_p<18.2$)} $z\ge 4.5$ quasars in the Southern hemisphere from {\refbf ten} objects prior to our observations to {\refbf 26} now.

We derive luminosities in two restframe bands, $M_{145}$ centred at 145~nm and $M_{300}$ at 300~nm, and show the impact of our work on the luminosity-redshift space of known quasars in Fig.~\ref{Lz}. For previously discovered quasars, authors most often listed $M_{145}$ as it is constrained by the $i$ and $z$ passbands in which the optical discovery is made, and usually also covered in the spectrum, so it can be determined while avoiding influence from the broad emission lines. Also, $M_{145}$ is the best available proxy for the ionising flux from the quasar. Measuring $M_{300}$, in contrast, requires $HK$ photometry, which has previously been available only for the most extreme objects that were bright enough to be detected by {\refbf the 2 Micron All-Sky Survey \citep[2MASS;][]{2MASS}}. $M_{300}$ is a better predictor of the bolometric luminosity of quasars as it is closer to the peak in their spectral energy distribution. {\refbf Now, the much deeper VISTA Hemisphere Survey (VHS) and the VIKING survey \citep{Edge13} are available, at least in the Southern hemisphere, and provide} very precise photometry for most quasars in our sample.

We estimate $M_{145}$ first for all objects in Table~\ref{tab_cands} with a simple linear interpolation of the SkyMapper $i$ and $z$~band photometry. For the subset of objects with spectra from the 2.3m-telescope, we also measure $M_{145}$ from the spectra after normalising them to the photometry. As we find no differences larger than $\pm 0.05$~mag, we report the simpler measure of $M_{145}$ that is available for all objects consistently. We do not report errors as they are smaller than the intrinsic long-term stochastic variability of quasars.

We measure $M_{300}$ in the best case from $H,K_s$ photometry in the VHS; {\refbf these objects have $H$ band errors up to $0.03$.} For a few objects, VHS currently has $J,K_s$ photometry but no $H$ band, and for these we estimate $H$ magnitudes from $J,K_s$ photometry. For most objects, we have all three near-infrared magnitudes, $J,H,K_s$, and for those we find that $(H-K_s)/(J-K_s)=0.41\pm 0.028$. We apply this tight relation to derive $H(J,K_s)$ and {\refbf combine} the measurement and relation errors; {\refbf the resulting $H$ band errors range from $0.03$ to $0.07$. Five} objects have no VHS coverage {\refbf but we find three of them in VIKING (J032444-291821, J221111-330245 and J230429-313426), one in UKIDSS (J090527+044342), and for the final one (J013539-212628)} we resort to {\refbf 2MASS photometry, which has a larger $H$ band error of $0.15$~mag. However, we stress that all these errors are smaller than the intrinsic long-term variability of quasars}.


\subsection{Radio detections}

None of the listed objects is detected in the 1.4 GHz VLA survey ''Faint Images of the Radio Sky at Twenty-Centimeters'' \citep[FIRST;][]{FIRST}, mostly due to its mostly Northern sky coverage. The 1.4 GHz NRAO-VLA Sky Survey \citep[NVSS;][]{NVSS} covers the sky at $\delta > -40\degr$ with a flux limit of 2.5~mJy; it contains four of the quasars; one of these, J052506.17-334305.6, is also detected by the 843~MHz Sydney University Molonglo Southern Survey \citep[SUMSS;][]{SUMSS}, which covers the polar half of the Southern hemisphere at $\delta < -30\degr$ with a limit of 5~mJy. Its fluxes of $183.8 \pm 5.9$ at 843 MHz and $188.3 \pm 5.7$ at 1.4 GHz suggest that it is a flat-spectrum radio quasar \citep[FSRQ;][]{UrryPadovani95}.

We use the radio-to-optical ratio $R=f_{\nu,\rm 6cm}/f_{\nu,B}$ defined by \citet{Kellermann89} to label objects as radio-loud ($R>10$) or radio-quiet ($R<1$); in that work the ratio is based on the flux density at 6~cm and 440~nm wavelength for a sample with average redshift of $\sim 0.5$, but for our objects these bands are redshifted on average ($z=4.75$) to $\sim 23$~cm, which is approximately the NVSS wavelength, and 1700~nm, which is in the NIR $H$ band. A sensitivity limit of $f<2.5$~mJy in the NVSS corresponds to $R<10$ for $H=16.6$, the typical magnitude of the faintest sources in this work, and $R<2$ for the $H=14.8$ of the brightest source. 

The four sources detected by NVSS and SUMSS range from $R\approx 40$ to $R\approx 300$ and are thus clearly radio-loud. Of these four, one is newly discovered and three have been known as quasars before. This is consistent with the fact that the fraction of radio-loud quasars is generally small, while known quasars in the Southern hemisphere have previously often been discovered by optical follow-up of radio source catalogues. All except one of the newly discovered quasars are thus radio-quiet.

\subsection{Gravitationally lensed quasars?}

Gravitationally lensed quasars are intrinsically very rare, but their fraction is higher among quasars of brighter apparent magnitude, as their brightness is boosted by the lensing. When quasars are lensed, they show multiple magnified images and hence their luminosity is overestimated. We thus look for signs of multiplicity among the objects in the {\it Gaia} catalogue, which is known to be incomplete at separations of $<2$ arcsec. We find just one object that has a {\it Gaia} neighbour within 5 arcsec, J090527.39+044342.3. There, we actually see two objects in the SkyMapper image coincident with two {\it Gaia} sources separated by 4 arcsec. The catalogued SkyMapper source is a blend of the two objects, with the centroid location close to the brighter of the two. The SDSS DR14\footnote{\url{https://skyserver.sdss.org/dr14/}} has separate photometry for the two sources and shows the fainter source to have colours that are different from the first and typical for a star. Hence, these are not expected to be two images of one object. The SkyMapper PSF magnitudes are 0.2 mag fainter than the SDSS measurements of the (brighter) quasar source; this is either a consequence of long-term variability, or the blending causes the SkyMapper PSF magnitude and the luminosity to be underestimated here. 



\begin{figure*}
\begin{center}
\includegraphics[angle=270,width=0.32\textwidth,clip=true]{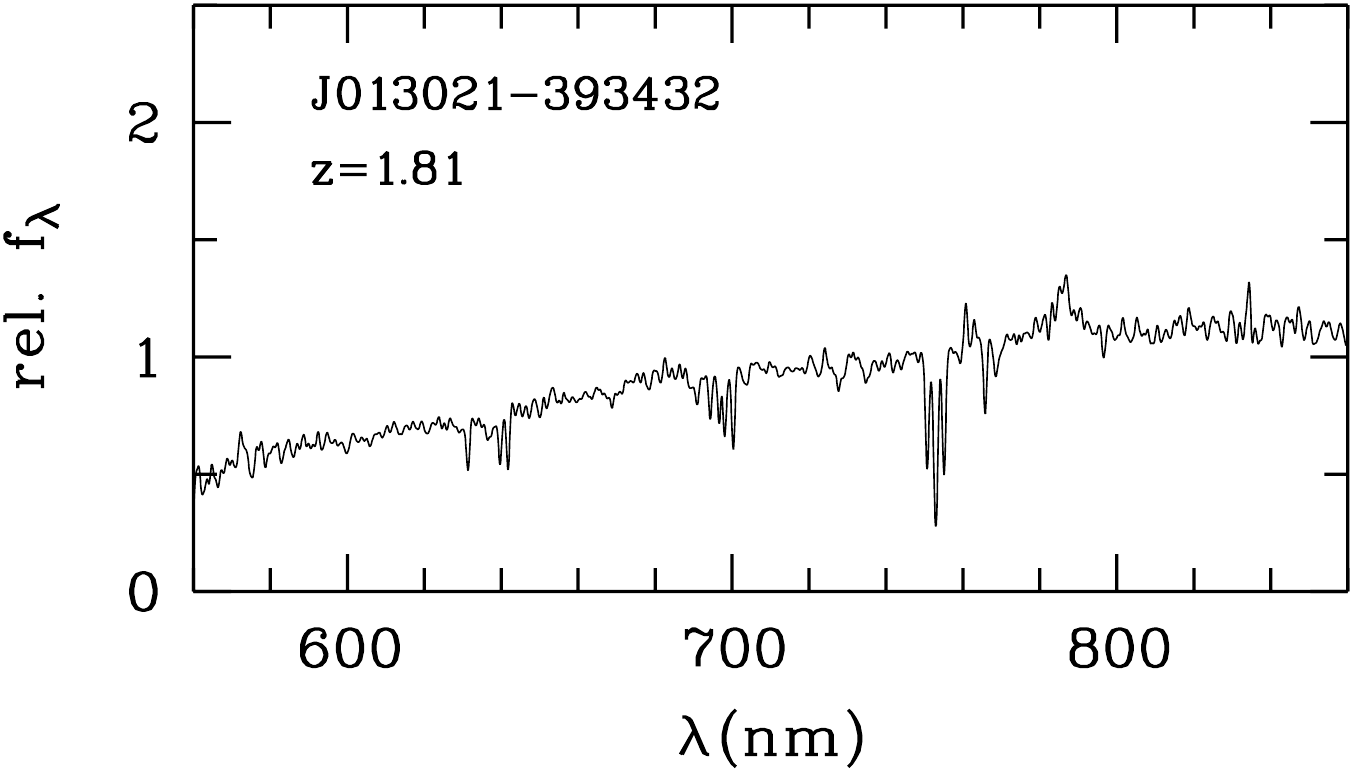}
\includegraphics[angle=270,width=0.32\textwidth,clip=true]{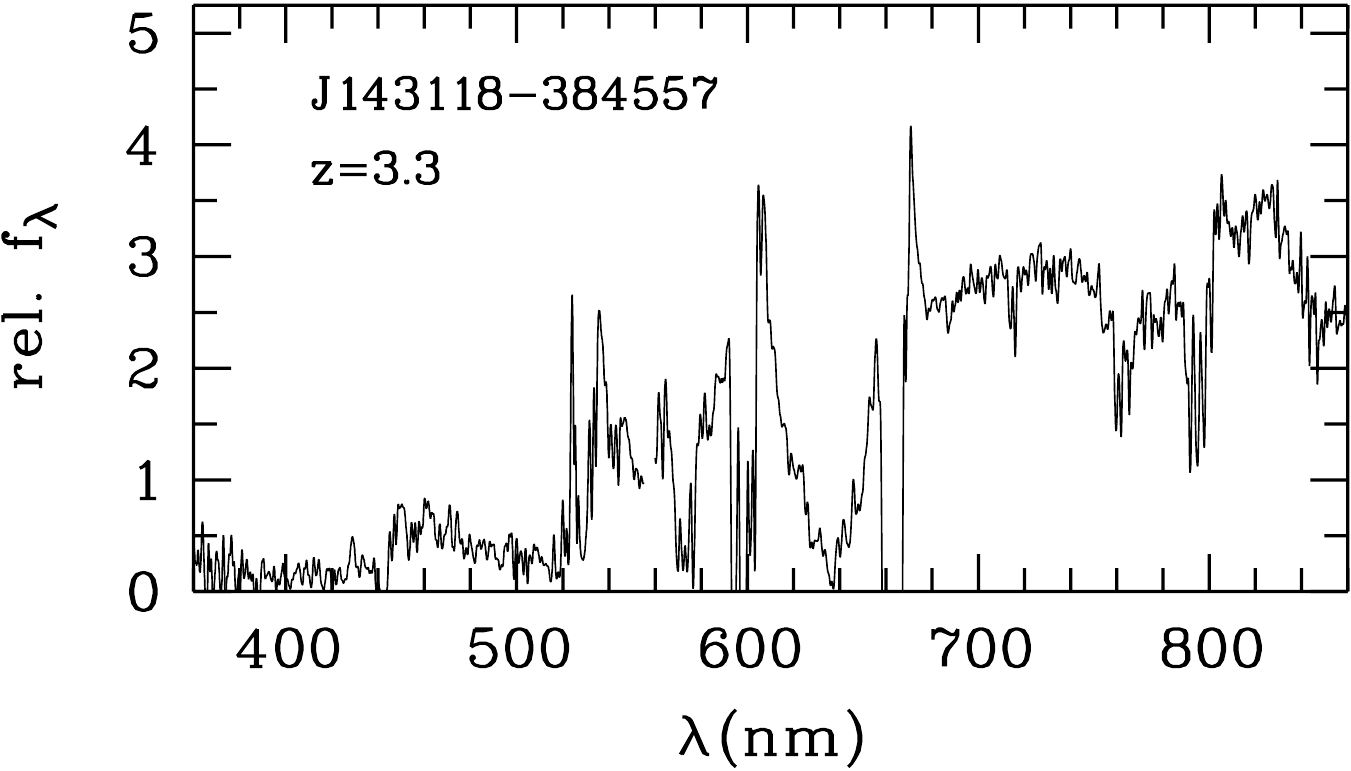}
\includegraphics[angle=270,width=0.32\textwidth,clip=true]{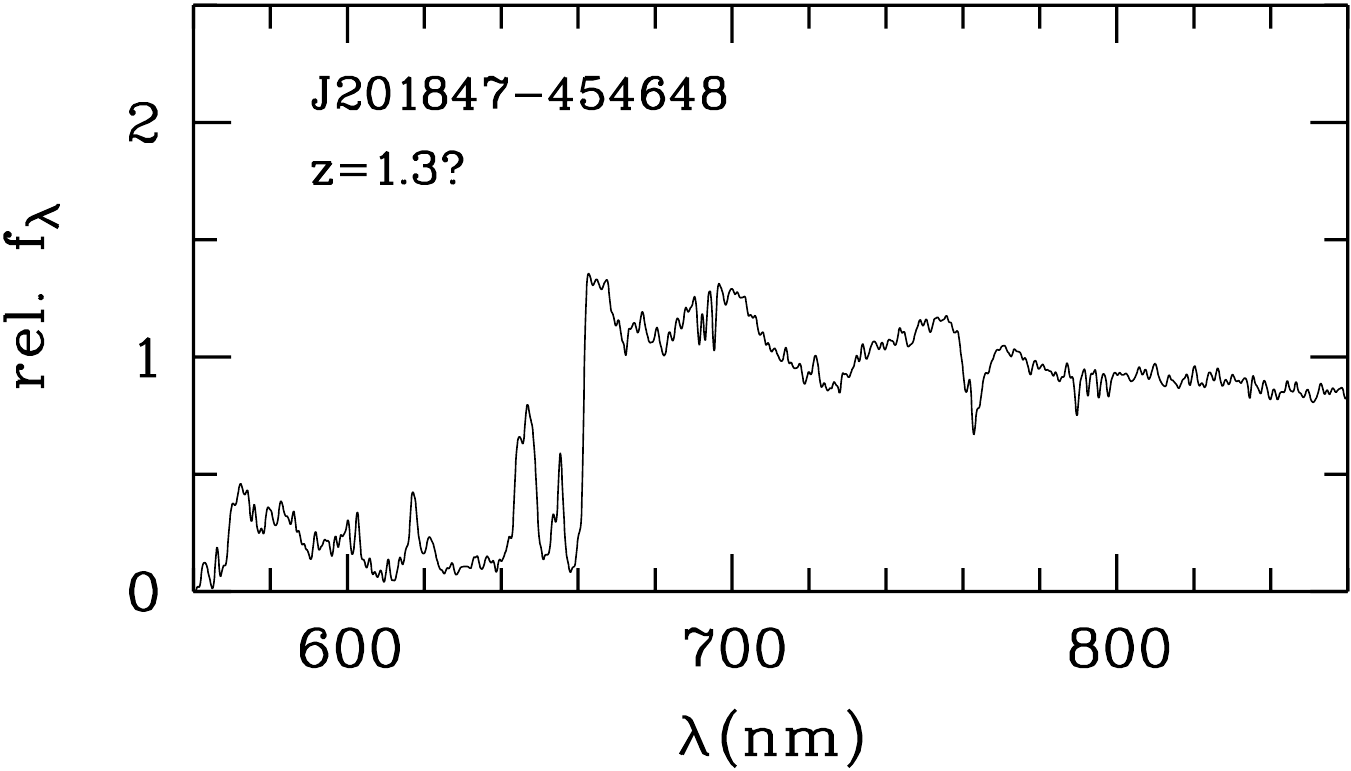}
\caption{Gallery of three lower-redshift quasar spectra. Left: a generally red continuum. Centre: a BALQSO. Right: an OFeLoBAL QSO.
\label{gallery_lozquasars}}
\end{center}
\end{figure*}

\subsection{Contaminants in candidate sample}

Most contaminants in our candidate sample turn out to be cool stars and include a few carbon stars and long-period variables. But we also identify {\refbf 24} quasars at lower redshift. Typical quasars at those lower redshifts are firmly excluded by our selection cuts, and the spectra reveal that these quasars are unusual. They fall into three categories (see Fig.~\ref{gallery_lozquasars} for example spectra):

\begin{enumerate}
    \item {\refbf Five} quasars at $z=0.5$ to $z=2$ have atypically red continua; \citet{Richards03} found that 6\% of quasars in a large SDSS sample fall into this category, which would be missed by surveys for quasars with typical, blue, colours.

    \item {\refbf Three} quasars at $z\approx 2.6$ to $3.3$ have strong broad absorption lines \citep[BALQSOs;][]{Weymann91, Scaringi09}, which render the emission lines nearly invisible. They are not easily separated from the other objects without resorting to spectra. \citet{Richards03} also find that BALs are more common among red quasars.
    
    \item {\refbf 16} objects have extremely wide absorption troughs at nearly full depth, such that it is very challenging to determine any redshift at all. Objects with qualitatively similar spectra have been found in the past and typed as OFeLoBALQSOs (Overlapping Iron Low-Ionisation Broad-Absorption line QSOs) by \citet{Hall02} and \citet{Zhang15}. Their spectra are believed to be nearly fully absorbed bluewards of the Mg~{\sc ii} line, which suggest a rough redshift based on the assumption that the unabsorbed continuum starts around 280~nm wavelength. If this is true, the objects we found would all reside at redshift $z\approx 1$ to $2$. This category of objects will be discussed in a separate publication.
\end{enumerate} 

{\refbf The number of lower-$z$ quasars with unusual SEDs is roughly as large as the number of true high-$z$ quasars and may come as a surprise when compared to earlier surveys. They also reach $\sim$1~mag brighter in $R_p$ than the $z>4.5$ quasars. It seems clear, how the lower-$z$ OFeLoBALQSOs are confused with $z>4.5$ quasars: the strong suppression bluewards of the Mg~{\sc ii} line simply mimics the effect of Lyman forest. But why are they more common than in previous searches? We believe, this is because we target high-$z$ quasars of such high and rare luminosity that we run out of objects, while the interlopers appear still with their natural abundance.}

\begin{figure}
\begin{center}
\includegraphics[angle=270,width=\columnwidth,clip=true]{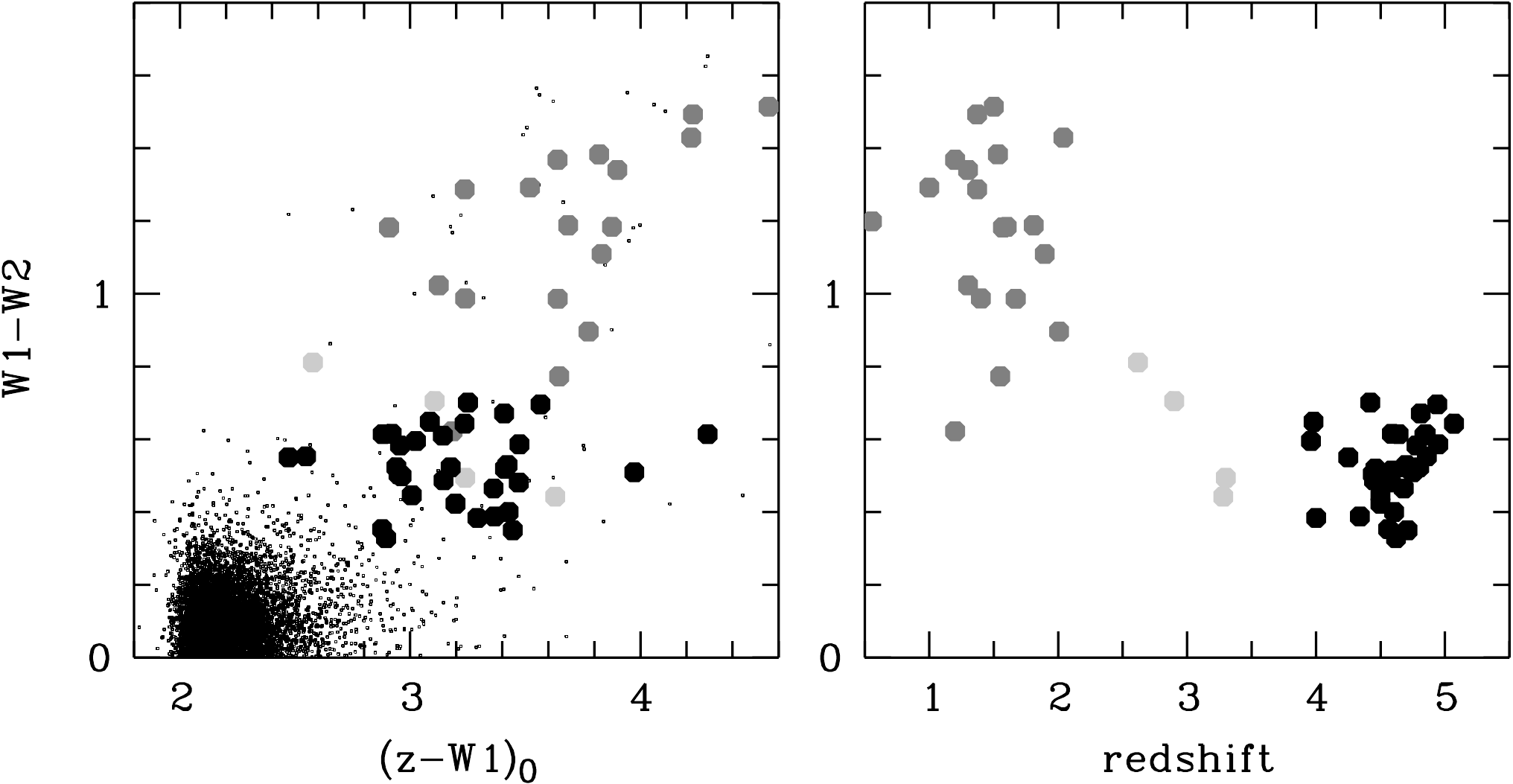} 
\caption{Confirmed quasars in our selected sample: at $z=[4,5]$ they have $W1-W2=[0.3,0.75]$, while lower-redshift quasars are redder. They are in our sample because of red continua, strong BAL features, or extreme absorption bluewards of the Mg~{\sc ii} line as in the case of OFeLoBAL QSOs.
\label{CCzDs}}
\end{center}
\end{figure}

We note that nearly all the $z<2.5$ quasars, whether those with generally red continua or the OFeLoBALQSOs with extreme absorption bluewards of the Mg~{\sc ii} line, have {\it WISE} colours of $W1-W2\approx [1,1.5]$, which are typical for quasars at $z<2$ (see Fig.~\ref{CCzDs}). In contrast, all of the high-redshift ($z\ge 4$) quasars {\refbf in this work have $W1-W2=[0.3,0.75]$ (see Fig.~\ref{CCzDs}), with a mean colour of $0.52$; the RMS scatter of $0.11$ appears to be mostly intrinsic, as the average error on the W1$-$W2 colour is $0.053$. The conclusion is, however,} that we could distinguish low-redshift interlopers from high-redshift objects using {\refbf appropriate} ranges in $W1-W2$ colour. We note that the high-redshift quasars reported in \citet{Wang16} include objects with $W1-W2$ colours as red as $\sim$1.7 despite moderate errors on the {\it WISE} photometry, although these are all fainter than our search regime. Either the errors of their {\it WISE} photometry is underestimated or high-redshift quasars with red mid-infrared colours are intrinsically rare at the highest luminosity studied here, so that none show up in our sample.

\section{Summary}

We used data from SkyMapper DR1 and DR2, from {\it Gaia} DR2 and from AllWISE to search for quasar candidates at $z>4.5$ in the Southern hemisphere. We focus on the bright end of the quasar distribution, where true quasars are outnumbered by cool stars to the most extreme degree possible. 

The role of {\it Gaia} DR2 is primarily to remove the vast majority of objects in the search space, which are cool stars in our own Milky Way that show noticeable parallax or proper motion. The {\it Gaia} data allow us to reduce the target density by a factor of {\refbf $\sim$40 now and several hundred in the future}. 

In this paper, we focus on very likely quasars and suppress stellar contamination further with a colour-based selection. In an area of $\sim${\refbf 12\,500}~deg$^2$, we find {\refbf 92 high-redshift candidates with a brightness} of $R_p<18.2$, among which were 13 already known high-redshift quasars. We {\refbf take spectra of} the others and {\refbf identify 21} more quasars, three of which have been identified and published independently since our observations.

This new sample {\refbf increases} the number of Southern $z\ge 4.5$ quasars at $R_p<18.2$ from {\refbf 10} to {\refbf 26}, and takes the number in the whole sky to {\refbf 42}. The {\refbf new} quasars are mostly in the luminosity range of $M_{145}\approx [-29,-28]$. At this point we refrain from investigating luminosity functions, because our search for more quasars will continue into more challenging parts of search space; those have been avoided in the past due to the abundance of cool stars that appear with similar colour to quasars and dramatically inflate the candidate samples. However, this area of search space is critical for assembling complete samples of quasars and will soon become accessible, owing to {\it Gaia} DR2 and because spectroscopic follow-up of many thousand targets at low sky density is becoming affordable with projects such as the Taipan Survey. 

{\refbf We also find among our candidates 24 lower-redshift quasars with unusual red spectra, most of which appear to be OFeLoBALQSOs. These are particularly prevalent in our sample relative to $z>4.5$ quasars, because we target a magnitude range where high-redshift quasars are truly rare, shifting the balance in favour of unusual low-$z$ objects. However, they can easily be distinguished using mid-infrared colours: while the rule $W1-W2<0.75$ retains all our high-redshift objects, it removes most low-redshift interlopers. }

In the future, we will venture into the stellar locus with massive spectroscopic follow-up, and revisit also part of the Northern sky with this approach, taking advantage of the {\it Gaia} motion measurements. We will also go deeper and to redshifts $z>5$. Finally, we will pursue measurements of black-hole masses for our sample.

\section*{Acknowledgements}

This research was conducted by the Australian Research Council Centre of Excellence for All-sky Astrophysics (CAASTRO), through project number CE110001020.
It has made use of data from the European Space Agency (ESA) mission {\it Gaia} (\url{https://www.cosmos.esa.int/gaia}), processed by the {\it Gaia} Data Processing and Analysis Consortium (DPAC, \url{https://www.cosmos.esa.int/web/gaia/dpac/consortium}). Funding for the DPAC has been provided by national institutions, in particular the institutions participating in the {\it Gaia} Multilateral Agreement.
The national facility capability for SkyMapper has been funded through ARC LIEF grant LE130100104 from the Australian Research Council, awarded to the University of Sydney, the Australian National University, Swinburne University of Technology, the University of Queensland, the University of Western Australia, the University of Melbourne, Curtin University of Technology, Monash University and the Australian Astronomical Observatory. SkyMapper is owned and operated by The Australian National University's Research School of Astronomy and Astrophysics. The survey data were processed and provided by the SkyMapper Team at ANU. The SkyMapper node of the All-Sky Virtual Observatory (ASVO) is hosted at the National Computational Infrastructure (NCI). Development and support the SkyMapper node of the ASVO has been funded in part by Astronomy Australia Limited (AAL) and the Australian Government through the Commonwealth's Education Investment Fund (EIF) and National Collaborative Research Infrastructure Strategy (NCRIS), particularly the National eResearch Collaboration Tools and Resources (NeCTAR) and the Australian National Data Service Projects (ANDS).
This work uses data products from the Wide-field Infrared Survey Explorer, which is a joint project of the University of California, Los Angeles, and the Jet Propulsion Laboratory/California Institute of Technology, funded by the National Aeronautics and Space Administration.
It uses data products from the Two Micron All Sky Survey, which is a joint project of the University of Massachusetts and the Infrared Processing and Analysis Center/California Institute of Technology, funded by the National Aeronautics and Space Administration and the National Science Foundation.
This paper uses data from the VISTA Hemisphere Survey ESO programme ID: 179.A-2010 (PI. McMahon).
This publication has made use of data from the VIKING survey from VISTA at the ESO Paranal Observatory, programme ID 179.A-2004. Data processing has been contributed by the VISTA Data Flow System at CASU, Cambridge and WFAU, Edinburgh.
The UKIDSS project is defined in Lawrence et al. 2007. UKIDSS uses the UKIRT Wide Field Camera (WFCAM; Casali et al. 2007) and a photometric system described in Hewett et al 2006. The pipeline processing and science archive are described in Irwin et al. (2008) and Hambly et al. (2008).
Support for this work was provided by NASA grant NN12AR55G. 
We thank Francis Zhong (University of Melbourne) for his contribution and ideas behind the final step of the script for extracting the spectrum from the 3D cube. We thank Ian McGreer for pointing out the SDSS photometry of J090527.39+044342.3 to us. {\refbf Finally, we thank an anonymous referee for comments improving the manuscript.}

\end{document}